\newcommand{\simgt}{\lower.5ex\hbox{$\; \buildrel > \over \sim \;$}}
\newcommand{\simlt}{\lower.5ex\hbox{$\; \buildrel < \over \sim \;$}}
\newcommand{\largesmall}{\lower.5ex\hbox{$\; \buildrel {\LARGE >} 
\over {\small <}$}}

\documentclass{ws-ijmpd}

\begin{document}

\markboth{Y.Habara and K.Yamamoto}
{Analytic Approach to Perturbed Einstein Ring with
Elliptical NFW Lens Model}

%
\catchline{}{}{}{}{}
%

\title{ANALYTIC APPROACH TO PERTURBED EINSTEIN RING WITH
ELLIPTICAL NFW LENS MODEL
}

\author{YUTA HABARA 
}

\address{Department of Physical Science, Hiroshima University,
Higashi-Hiroshima 739-8526,Japan
\\
habara@theo.phys.sci.hiroshima-u.ac.jp
}

\author{KAZUHIRO YAMAMOTO}

\address{Department of Physical Science, Hiroshima University,
Higashi-Hiroshima 739-8526,Japan\\
kazuhiro@hiroshima-u.ac.jp
}

\maketitle


\begin{abstract}
We investigate the validity of the approximate method to describe
a strong gravitational lensing which was extended by 
Alard on the basis of a perturbative approach to an Einstein ring. 
Adopting an elliptical Navarro-Frenk-White (NFW) lens model, 
we demonstrate how the approximate method works, focusing on 
the shape of the image, the magnification, caustics, 
and the critical line. Simplicity of the approximate method 
enables us to investigate the lensing phenomena in an analytic way. 
We derive simple approximate formulas which characterise 
a lens system near the Einstein ring. 

\keywords{Gravitational lens}
\end{abstract}


\section{Introduction}
Cold dark matter is one of the most important components
in the universe. The cosmic microwave background anisotropies 
and the large scale distribution of galaxies cannot be 
naturally explained without the cold dark matter component.
The mean density parameter of the cold dark matter has been 
measured precisely,\cite{WMAP} but its true character
has not been identified.
The elementary particle physics predicts possible candidates 
of the cold dark matter, and many experiments are ongoing 
aiming at a direct measurement. 

The cold dark matter is considered to be distributed 
associated with each galaxy, forming dark matter halo.
Then, the investigation of the structure of the halos
is quite important in exploring the nature and the origin
of the dark matter. 
The strong gravitational lensing is a useful probe of the 
halo-structure (see, e.g.,\cite{ZR} for a review). 
Especially, a lens system near Einstein ring is useful 
because a wealth of information can be obtained.\cite{Kochanek} 

Besides, the strong lensing systems are also useful as a tool of the 
dark energy study.\cite{FutamaseII}\cdash\cite{Futamase}
Because of the recent observational developments, many strong
lensing systems have been found. The strong lensing statistics
is now becoming one of the powerful tool for 
exploring the nature of the dark energy.\cite{Oguri}
Future dark energy surveys will detect much more strong lensing 
systems (see, e.g.,\cite{LSST}), and the strong lensing system 
will play a more important roll in cosmology.

In realistic situations, the mass distribution in a halo is not 
simple, which makes reconstruction of the lens model complicated.
The lens equation is complicated for a non-spherical lens model, 
which  needs to be solved numerically. 
Then, analytic approximate approach to strong lensing system is useful, 
if its validity and accuracy are guaranteed. 
A perturbative approach to the lensing system close to the Einstein 
ring configuration was developed, e.g.,\cite{Blandford,SEF}. 
Recently, Alard extended the perturbative approach, which is
applied to analyse lensing systems.\cite{Alard07}\cdash\cite{Alardnew} 

In the present paper, we investigate the validity of the perturbative 
approach to the lensing system close to an Einstein ring, assuming 
an elliptical lens model. We demonstrate the validity of the 
perturbative approach quantitatively, by comparing with an exact 
approach on the basis of the numerical method, focusing on the shape 
of the image, the magnification, the caustics, and the critical line.
Using the approximate method, expanded in terms of the ellipticity
parameter of the lens model, we derive simple 
approximate formulas which characterise an elliptical lensing system 
near the Einstein ring in an analytic way.

This paper is organized as follows: In section 2, we briefly review
the basic formulas for the gravitational lensing and the perturbative
approach to a perturbed Einstein ring, based on the work by
Alard.\cite{Alard07,Alard08} In section 3, we compare the perturbative
approach with the exact approach that relies on a numerical method,
focusing on the shape of lensed images, the caustics, the critical
curve, and the magnification, respectively. We demonstrate the validity of the 
perturbative approach at a quantitative level. 
In section 4, some useful formulas are presented, which are derived
using the perturbative approach in the analytic manner. 
Section 5 is devoted to summary and conclusions. 
Throughout the paper, we use the unit in which the speed of 
light equals 1.

\section{Basic Formulas}
\subsection{General basis}
We briefly review basic formulas for the strong lensing (e.g.,\cite{SEF}). 
The deflection angle of a lens object is determined by 
\begin{eqnarray}
&&\vec{\hat{\alpha}}= 2\int_0^{\chi_{\rm S}} 
{\partial \Phi({\vec\xi,\chi})\over \partial \vec\xi} d\chi,
\end{eqnarray}
where $\Phi$ is the gravitational potential of the lens object, 
$\chi$ is the radial coordinate connecting the observer and
the lens object, $\vec\xi$ is the two dimensional vector 
on the lens plane, which is orthogonal to the coordinate $\chi$. 
The gravitational potential $\Phi$ is related to the mass density 
distribution of the lens $\rho({\vec\xi,\chi})$ by 
the Poisson equation,
\begin{eqnarray}
&&\triangle \Phi({\vec\xi,\chi})=4\pi G \rho({{\vec\xi,\chi}}),
\end{eqnarray}
where $\triangle $ denotes the 3-dimensional Laplacian, and 
$G$ is the gravitational constant.

Introducing the surface mass density $\Sigma(\vec\xi)$, 
which is the projected mass density on the lens plane, 
\begin{eqnarray}
&&\Sigma(\vec\xi)=\int_0^{\chi_{\rm S}} \rho({{\vec\xi,\chi}})d\chi,
\end{eqnarray}
and the lensing potential, 
\begin{eqnarray}
&&\psi(\vec\xi)=\int_0^{\chi_{\rm S}} \Phi({{\vec\xi,\chi}})d\chi,
\end{eqnarray}
which are related by 
\begin{eqnarray}
&&\triangle^{(2)}\psi(\vec\xi)=4\pi G\Sigma(\vec\xi),
\end{eqnarray}
where  $\triangle^{(2)} $ denotes the 2-dimensional Laplacian.
The solution is
\begin{eqnarray}
\psi(\vec\xi)=2G \int d^2\xi' \Sigma(\vec\xi')
\log|\vec\xi-\vec\xi'|+
{\rm constant}.
\end{eqnarray}
Then, the deflection angle is 
\begin{eqnarray}
\vec{\hat{\alpha}}&=& 2
{\partial \psi({\vec \xi})\over \partial \vec\xi}
=
 4 G\int d^2\xi' \Sigma(\vec\xi')
{\vec\xi-\vec\xi'\over | \vec\xi-\vec\xi'|^2}.
\end{eqnarray}

Now we consider the gravitational lens equation,
%
\begin{eqnarray}
\vec\eta={D_{\rm S}\over D_{\rm L}}\vec\xi-{D_{\rm LS}}\vec{\hat\alpha}, 
\end{eqnarray}
where $D_{\rm LS}$ is the angular diameter distance between 
the lens and a source object, $D_{\rm S}$ is the distance
between the observer and the source, $D_{\rm L}$ is the distance 
between the observer and the lens, and $\vec\eta$ is the two
dimensional vector on the source plane orthogonal to the 
coordinate $\chi$.
Introducing a characteristic length in the lens plane, $\xi_0$, 
and $\eta_0=\xi_0 D_{\rm S}/D_{\rm L}$ in the source plane, 
we define
\begin{eqnarray}
&&\vec y= \vec\eta/\eta_0
\\
&&\vec x=\vec \xi/\xi_0,
\end{eqnarray}
then, the lens equation becomes
\begin{eqnarray}
\vec y=\vec x-\nabla_{\vec x} \phi({\vec x}),
\label{lenseq}
\end{eqnarray}
where we defined
\begin{eqnarray}
\phi({\vec x})={1\over \pi} \int d^2 x' \kappa(\vec x')\log |\vec x-\vec x'|
\end{eqnarray}
with
\begin{eqnarray}
&&\kappa(\vec x)= {\Sigma(\vec\xi)\over \Sigma_{\rm cr}},
\\
&&\Sigma_{\rm cr}^{-1}={4\pi G D_{\rm LS}D_{\rm L}\over D_{\rm S} }.
\end{eqnarray}

\subsection{Perturbative approach to Einstein ring}
Next, we review the perturbative approach to the Einstein ring developed 
by Alard \cite{Alard07,Alard08} (cf. \cite{Blandford,SEF}). 
When the projected density 
of the lens $\Sigma$ is circularly symmetric and the source
is located at the origin of the source plane, ${\vec y}=0$, 
an Einstein ring is formed. 
The radius of the Einstein ring is determined by 
\begin{eqnarray}
\vec x-{\partial \phi_0(|{\vec x}|)\over \partial \vec x}=0,
\label{Ring}
\end{eqnarray}
where $\phi_0(|\vec x|)$ denotes the circularly symmetric lens potential.
We denote the solution of Eq.~(\ref{Ring}) by $\vec x=\vec x_E$. 
Thus, $|\vec x_E|$ is the Einstein radius. Hereafter, we use
the notation $x=|\vec x|$ and $x_E=|\vec x_E|$.

We consider the perturbative approach to the Einstein ring, 
then assume that the deviation from the circularly 
symmetric lens is small. Introducing the small deviation, 
which are denoted by the quantities with $\delta$,  
\begin{eqnarray}
&&\vec y = \delta \vec y,
\\
&&\vec x = \vec x_E+\delta \vec x,
\\
&&\phi(\vec x)=\phi_0(x)+\delta \phi({\vec x}),
\end{eqnarray}
the lens equation (\ref{lenseq}) is rephrased as
\begin{eqnarray}
\delta \vec y ={\vec x}_E+\delta \vec x-\nabla_{\vec x} [
\phi_0(x)+\delta\phi({\vec x})]\Big|_{\vec x=\vec x_E+\delta \vec x}.
\end{eqnarray}
Assuming that the deviation from the circularly symmetric lens 
is small, we introduce the small expansion parameter 
$\varepsilon$. Explicitly, we assume 
\begin{eqnarray}
&&\delta \vec y = {\cal O}(\varepsilon),
\\
&&\delta \vec x = {\cal O}(\varepsilon),
\\
&&\delta \phi({\vec x})= {\cal O}(\varepsilon).
\end{eqnarray}
We find the following lens equation at the lowest order of $\varepsilon$,
\begin{eqnarray}
\delta \vec y =\delta \vec x
-[(\delta \vec x\cdot\nabla_{\vec x})\nabla_{\vec x}\phi_0(x)
+\nabla_{\vec x}\delta\phi({\vec x})]\Big|_{\vec x=\vec x_E},
\label{lenseqp}
\end{eqnarray}
where we used Eq.~(\ref{Ring}). 

We consider a circular source with the radius $\delta r_s$, 
whose centre is located at the coordinate $(\delta y_{10},\delta y_{20})$ 
on the source plane. Then, the circumferences of the source is 
parameterised as 
\begin{eqnarray}
  &&\delta y_1=\delta y_{10}+\delta r_s \cos\varphi 
\label{sourcea} 
\\ 
&&\delta y_2=\delta y_{20}+\delta r_s \sin\varphi
\label{sourceb}
\end{eqnarray}
with the parameter $\varphi$ in the range $0\leq \varphi \leq 2\pi$, 
where we assume that 
$\delta y_{10}$, $\delta y_{20}$ and $\delta r_s$ are 
the quantities of the order of $\varepsilon$.
Similarly, we may rewrite the image position 
$\vec x = \vec x_E+\delta \vec x$ as \cite{Alard07}
\begin{eqnarray}
  &&x_1=(x_E+\delta x)\cos\theta,
\label{x1dd}
\\
  &&x_2=(x_E+\delta x)\sin\theta.
\label{x2dd}
\end{eqnarray}
Here, we don't need to consider the perturbation of $\theta$
because of the symmetry of the un-perturbed image.\cite{Alard07} 
In the un-perturbed situation, the image 
of a point source is a circle, and there are an infinite 
number of the image at any  $\theta$.
Then, at any angular position of the perturbed point, there
is always an un-perturbed point at the same angular position 
on the circle.


The lens equation (\ref{lenseqp}) yields 
\begin{eqnarray}
  &&\delta y_{10}+\delta r_s \cos\varphi=\bigl[\delta x\cos\theta\left(
  1-\partial_x^2\phi_0(x)\right)
\nonumber
\\&&~~~~~
  -\cos\theta \partial_x \delta\phi(x,\theta)+{\sin\theta\over x}
   \partial_\theta \delta\phi(x,\theta)\bigr]\Big|_{x=x_E},
\label{leneqptx}
\\
  &&\delta y_{20}+\delta r_s \sin\varphi=\bigl[\delta x\sin\theta\left(
  1-\partial_x^2\phi_0(x)\right)
\nonumber
\\&&~~~~~
  -\sin\theta \partial_x \delta\phi(x,\theta)-{\cos\theta\over x}
   \partial_\theta \delta\phi(x,\theta)\bigr]\Big|_{x=x_E}.
\label{leneqpty}
\end{eqnarray}
Combining these equations, we have
\begin{eqnarray}
  &&\delta x={1\over (1-\partial_{x}^2 \phi_0(x))}\Bigl[
\partial_{x}\delta\phi(x,\theta)+\delta y_{10}\cos\theta
+\delta y_{20} \sin\theta
\pm \sqrt{\Delta^2(x,\theta)}\Bigr]\Bigr|_{x=x_E},
\nonumber
\\&&~~~~~~~
\label{dreq}
\end{eqnarray}
where we defined
\begin{eqnarray}
  &&\Delta^2(x,\theta)=\delta r_s^2
-\Bigl(
  {1\over x}\partial_\theta\delta \phi(x,\theta)
  -\delta y_{10}\sin\theta+\delta y_{20} \cos\theta\Bigr)^2.
\label{Delta2}
\end{eqnarray}
This is the formula derived by Alard.\cite{Alard07,Alard08}

For comparison, we summarised the corresponding formula
without the perturbative approximation in the appendix A. 
One can derive Eq.~(\ref{dreq}) from Eq.~(\ref{exactimage}).

\subsection{Perturbative NFW lens model}
A mass model commonly used for strong lensing is based on 
high-resolution numerical simulations of dark-matter halos 
in the $\Lambda$CDM framework by Navarro, Frenk and White (1996, 
1997; hereafter NFW\cite{NFWI,NFW2}), in which the density profile 
is parameterised by the scale radius $r_s$ and the constant 
$\rho_s$, 
\begin{eqnarray}
\rho(R)=\frac{\rho_{s}}{\left(R/r_s\right)\left(1+R/r_s\right)^2},
\end{eqnarray}
where $R=\sqrt{\xi^2+\chi^2}$ is the 3-dimensional length. 
Choosing $r_s=\xi_0$, the lens potential becomes \cite{Bartelmann}
\begin{eqnarray}
\phi_{\rm NFW}(x)=4\kappa_sF(x),
\end{eqnarray}
where we defined 
\begin{eqnarray}
F(x)=
{1\over 2}\log^2\frac{x}{2}+\left\{
\begin{array}{cc}
+2\arctan^2\sqrt{\frac{x-1}{x+1}} & (x>1) \\
-2{\rm arctanh}^2 \sqrt{\frac{1-x}{1+x}} & (x<1) \\
\log x & (x=1)
\end{array}
\right.
\label{defNFWphi}
\end{eqnarray}
and
\begin{eqnarray}
\kappa_s={\rho_s \xi_0 \over \Sigma_{\rm cr}}.
\end{eqnarray}
We denote the solution of the lens equation for 
the circular NFW lens model by $u_0$, which satisfies
\begin{eqnarray}
 u_0= {\partial \over \partial u_0}\phi_{\rm NFW}(u_0).
\label{ERNFW}
\end{eqnarray}

In the present paper, for an asymmetric lens model, we adopt 
the potential 
\begin{eqnarray}
\phi({\vec x})=\phi_{\rm NFW}\left(x\sqrt{1-\eta\cos2\theta}\right).
\end{eqnarray}
Instead of $\vec x$ and $\vec y$, we introduce $\vec{\widetilde x}$
and $\vec{\widetilde y}$, which is normalised by the 
Einstein radius $u_0$ defined with Eq.~(\ref{ERNFW}),
\begin{eqnarray}
&& \vec x= u_0 \vec{\widetilde x},
\\
&& \vec y= u_0 \vec{\widetilde y}.
\end{eqnarray}
Then, the lens equation is rewritten
\begin{eqnarray}
  \vec{\widetilde y}=\vec{\widetilde x}-
  \nabla_{\vec{\tilde x}}{\widetilde \phi}
\end{eqnarray}
with
\begin{eqnarray}
{\widetilde \phi}=
 {4\kappa_s\over u_0^2} 
 F\left(u_0{\widetilde x}\sqrt{1-\eta\cos2\theta}
\right),
\end{eqnarray}
where ${\widetilde x}=|\vec{\widetilde x}|$, 
$F(x)$ is defined by Eq.~(\ref{defNFWphi}),
and \begin{eqnarray}
  &&\left({4\kappa_s\over u_0^2}\right)^{-1}=
\log\frac{u_0}{2}+
\left\{
\begin{array}{cc}
{2\over \sqrt{u_0^2-1}}\arctan\sqrt{\frac{u_0-1}{u_0+1}} & (u_0>1) \\
{2\over \sqrt{1-u_0^2}}{\rm arctanh}\sqrt{\frac{1-u_0}{1+u_0}} & (u_0<1) \\
1 & (u_0=1)
\end{array}
\right.
\end{eqnarray}
Finally, the potential of the elliptical NFW 
lens is written as
\begin{eqnarray}
\widetilde \phi= \widetilde \phi_0( \widetilde x)
+ \widetilde{\delta\phi}( \widetilde x,\theta)
\label{insummary}
\end{eqnarray}
with
\begin{eqnarray}
 && \widetilde \phi_0( \widetilde x)= {4\kappa_s\over u_0^2} 
 F(u_0 \widetilde x),
\\
 && \widetilde{\delta\phi}( \widetilde x,\theta)
={4\kappa_s\over u_0^2} 
\Bigl(F\bigl(u_0 \widetilde x\sqrt{1-\eta\cos2\theta}
\bigr)
-F(u_0 \widetilde x)\Bigr).
\label{deltaphiNFW}
\end{eqnarray}
In the appendixes A and B, useful formulas related with the
elliptical NFW lens potential are summarized. In these appendixes, 
only the case $u_0\widetilde x\sqrt{1-\eta\cos2\theta}<1$ is 
described, but the case $u_0\widetilde x\sqrt{1-\eta\cos2\theta}>1$ 
is obtained by the analytic continuation.

\section{Validity of the Perturbative approach}
We here investigate the validity of the perturbative approach, 
comparing with results without any approximation. 
For being definite, we consider the following three cases. 

(a) Exact approach without any approximation. 

(b) Perturbative approach described in the previous section. 

(c) Approximate approach: the perturbative approach (b) plus the 
lowest-order expansion of $\eta$ (See also the appendix C).

\subsection{Image}
We consider the lensed image of the circumference of the circular 
source, whose center is located at $(\widetilde{\delta y_{10}},
\widetilde{\delta y_{20}})$. The source's radius is $\widetilde{\delta r_s}$. 
In the exact approach (a), the circumference of the lensed image
is obtained by solving Eq.~(\ref{exactimage}), which is derived 
in the appendix A. Eq.~(\ref{exactimage}) can be solved
with an iterative method numerically.
On the other hand, in the perturbative approach (b), 
the circumference follows Eq.~(\ref{dreq}), which is equivalent to
\begin{eqnarray}
  &&\widetilde {\delta x}={1\over (1-\partial_{\tilde x}^2 
\widetilde{\phi_0}(\widetilde x))}\Bigl[
\partial_{\widetilde x}\widetilde{\delta\phi}(\widetilde x,\theta)+\widetilde{\delta y_{10}}
\cos\theta
+\widetilde{\delta y_{20}} \sin\theta
\pm \sqrt{\widetilde\Delta^2(\widetilde x,\theta)}\Bigr]\Bigr|_{\widetilde x=1},
\nonumber
\\&&~~~~~~~
\label{dreqap}
\end{eqnarray}
with 
\begin{eqnarray}
  &&\widetilde\Delta^2(\widetilde x,\theta)=\widetilde{\delta r_s}^2
-\Bigl(
  {1\over \widetilde x}\partial_\theta\widetilde{\delta \phi}
(\widetilde x,\theta)
  -\widetilde{\delta y_{10}}\sin\theta
  +\widetilde{\delta y_{20}}\cos\theta \Bigr)^2.
\label{Delta2ap}
\end{eqnarray}
Eqs.~(\ref{dreqap}) and (\ref{Delta2ap}) are the same as
Eq.~(12) in the reference Alard \cite{Alard07}.\footnote{
It is useful to summarize the differences of the notation between 
the present paper and the reference by Alard \cite{Alard07}.
Our equations can be obtained by the following transformation 
from the equations in the paper by Alard \cite{Alard07}, 
$r\rightarrow\widetilde x$, $\mathrm{d}r\rightarrow\widetilde {\delta x}$,
$\phi_0\rightarrow\widetilde{\phi_0}$,
$\psi\rightarrow\widetilde{\delta\phi}$, 
$R_0\rightarrow\widetilde{\delta r_s}$,
$x_0\rightarrow\widetilde{\delta y_{10}}$,
and $y_0\rightarrow\widetilde{\delta y_{20}}$.}
In the approximate approach (b), we use 
Eqs.~(\ref{dxxphi})$\sim$(\ref{dthetadeltaphi}) to find the solution,
Eq.~(\ref{dreqap}).
In the approximate approach (c), we use the quantities of the 
lowest-order of $\eta$, given by
Eqs.~(\ref{dxdeltaphiap}) and (\ref{dthetadeltaphiap}), 
instead of Eqs.~(\ref{dxdeltaphi}) and (\ref{dthetadeltaphi}).

\begin{figure}[hptb]
\begin{center}
\begin{tabular}{cc}
\includegraphics[width=61mm, height=61mm]{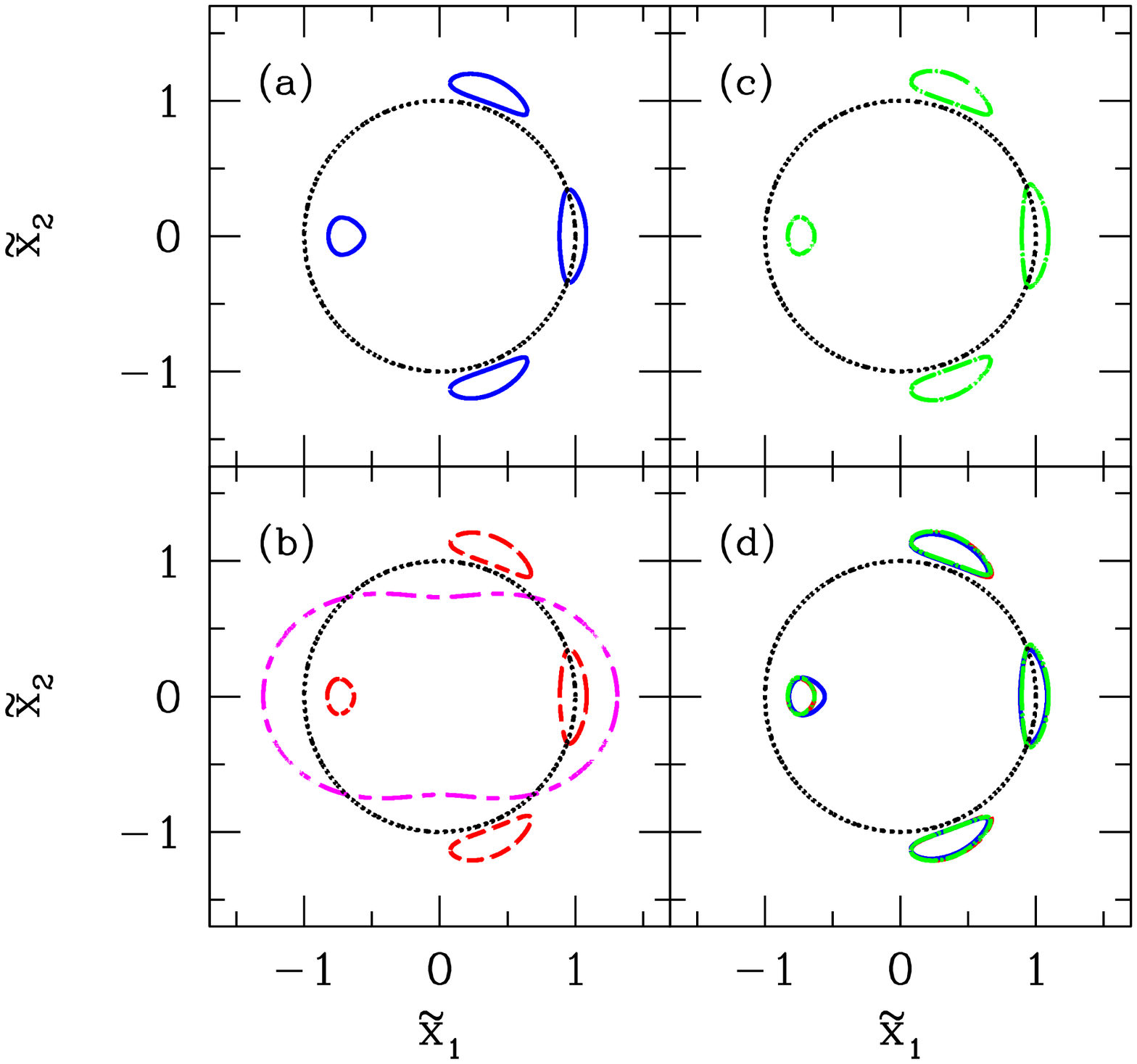}
&
\includegraphics[width=61mm, height=61mm]{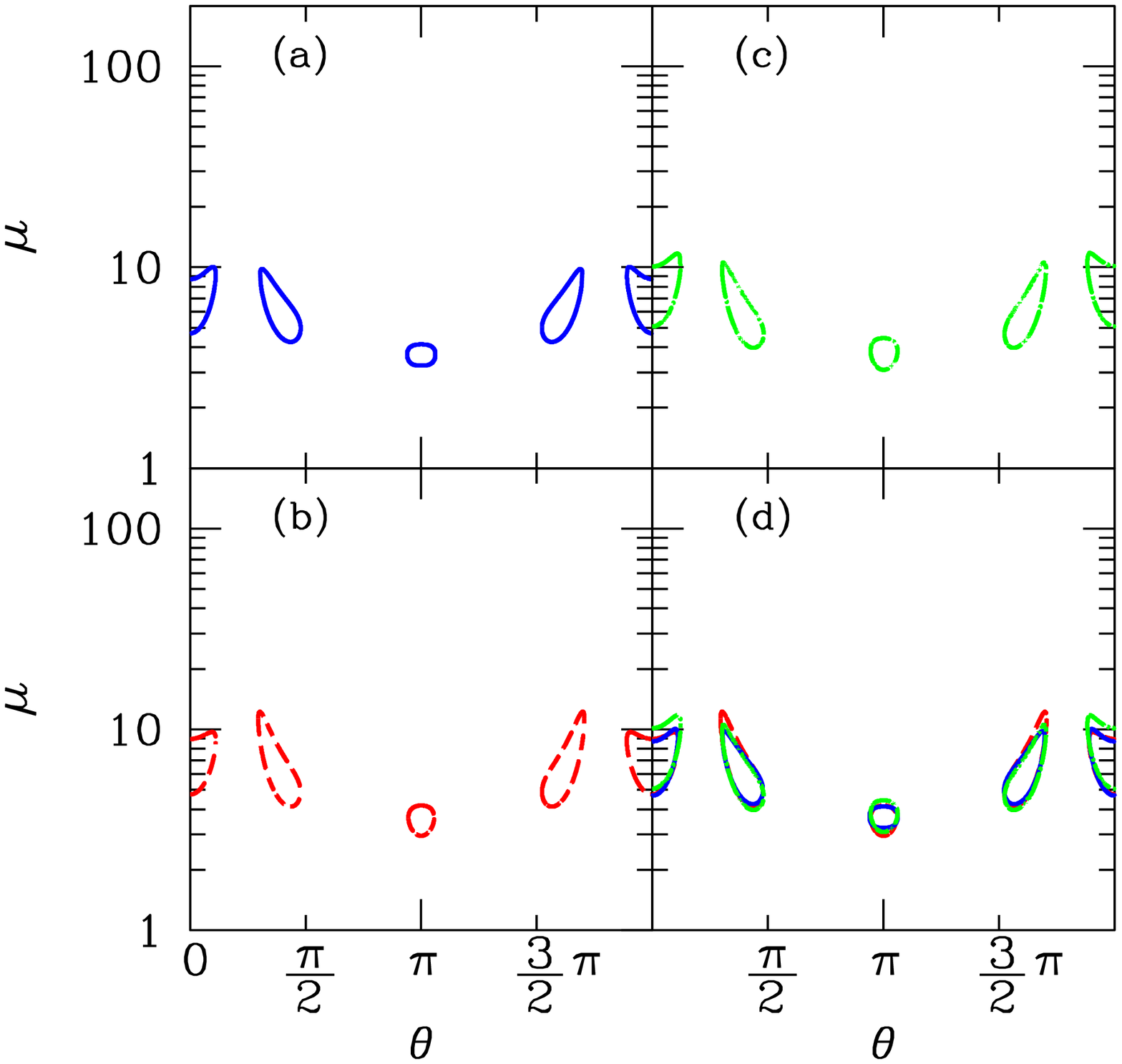}
\\
\includegraphics[width=61mm, height=61mm]{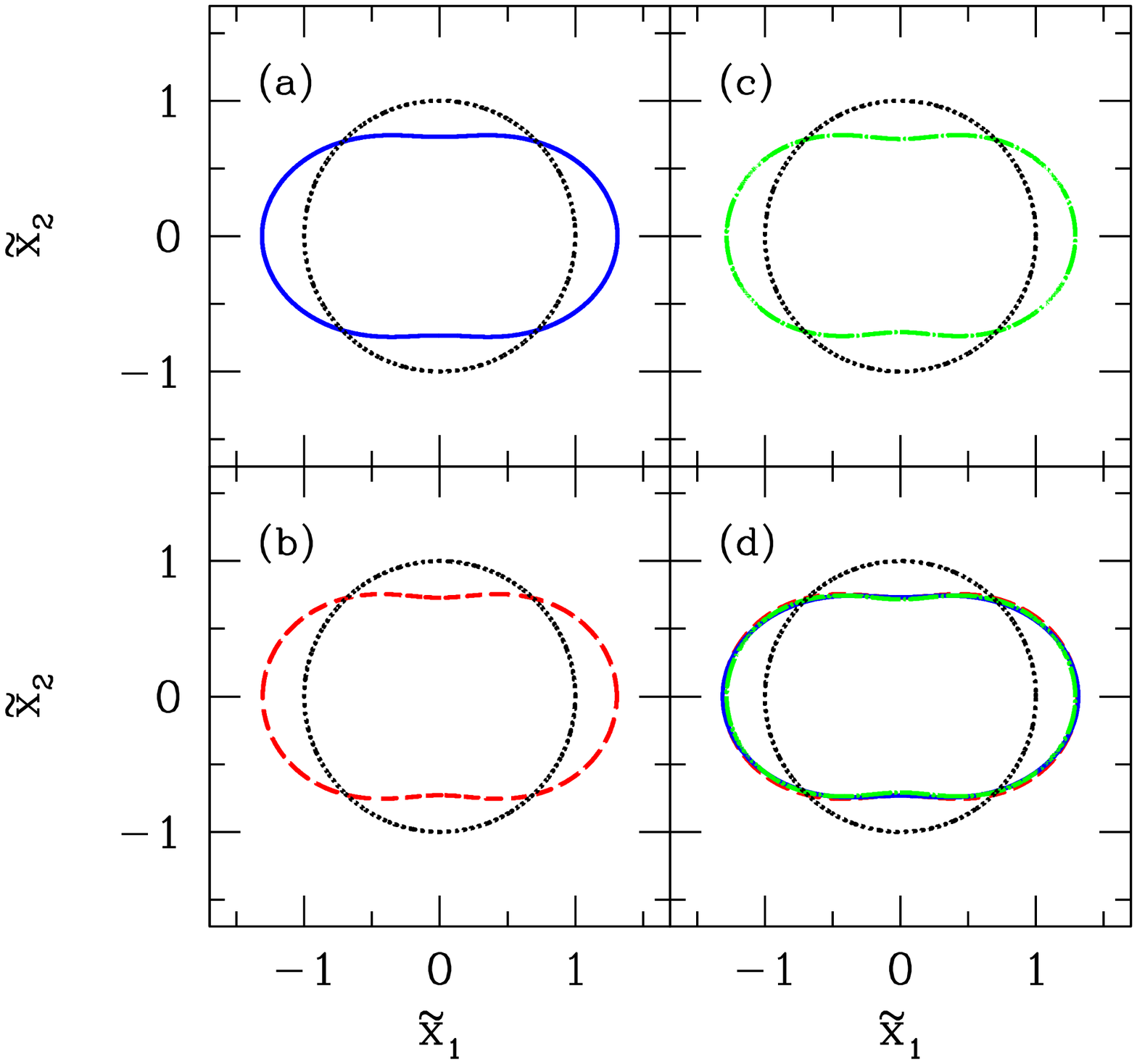}
&
\includegraphics[width=61mm, height=61mm]{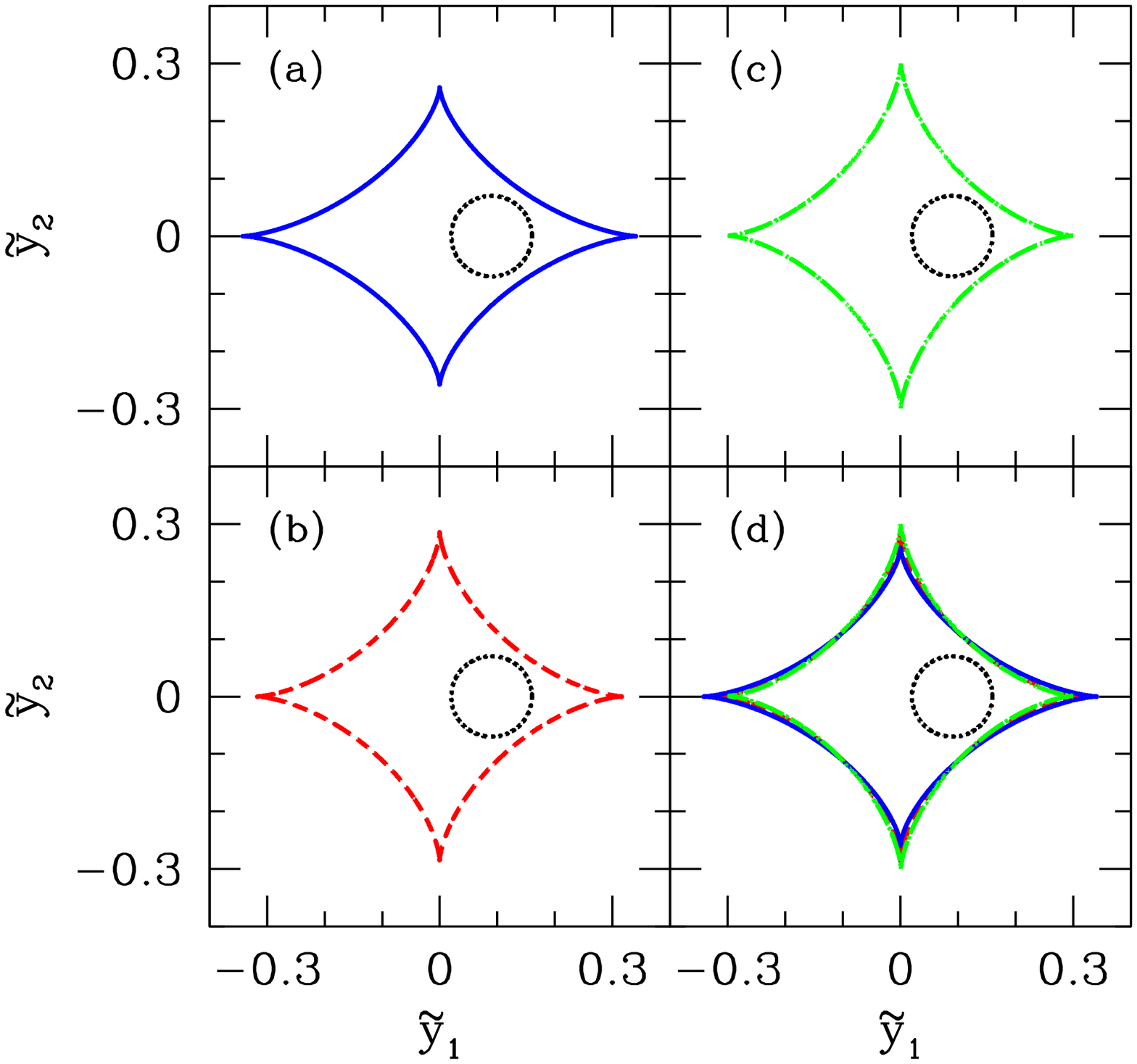}
\end{tabular}
\end{center}
\caption{ Comparison of the three approaches, 
(a) exact approach, (b) perturbative approach and (c) approximate approach 
of the lowest-order of the expansion of $\eta$. 
The panel labelled by (a), (b) and (c) corresponds to the approach (a), 
(b) and (c), respectively, while (d) plots all the three approaches 
for comparison. 
The upper left panels show the lensed image
of the circumference of a circular source. The dotted circle is the
Einstein ring. The long and short dashed curve is the critical line.  
The parameters $\widetilde {\delta r_s}=0.07$, $\widetilde {\delta y_{10}}=0.09$,
$\widetilde {\delta y_{20}}=0$, $\eta=0.15$ and $u_0=0.5$.
This is type I.
The upper right panels show the magnification factor $\mu$
on the circumference of the images 
as a function of the angular coordinate of the image plane.
The lower left panels show the critical line. 
The dotted circles are the Einstein ring.
The lower right panels show the caustics on the 
lens plane. The circle is the circumference of source.
}
\label{figI}
\end{figure}

The upper left panels of Fig.~\ref{figI} show
the lensed image. (a) is the exact approach, 
(b) is the perturbative approach, and (c) 
is the approximate approach, respectively.
The panel (d) plots these three approaches for comparison. 
Here we adopted the parameters $\widetilde {\delta r_s}=0.07$, 
$\widetilde{\delta y_{10}}=0.09$, $\widetilde {\delta y_{20}}=0$, $\eta=0.15$ 
and $u_0=0.5$. In this case, the lens effect splits the image 
into four. The dashed circle in each panel is the Einstein radius 
of a point source. 

\subsection{Magnification factor}
The magnification factor due to the gravitational lensing 
is given by the inverse of the determinant of the Jacobian 
matrix,\cite{SEF}
\begin{eqnarray}
  \mu=|J|^{-1}=\biggl|{\rm det}{\partial \vec y\over \partial \vec x}\biggr|^{-1}.
\end{eqnarray}
For a general lens potential, 
the determinant of the Jacobian matrix can be written as 
\begin{eqnarray}
&&J={1\over \widetilde x} \biggl[\Bigl(1-{\partial^2 \widetilde\phi\over 
\partial {\tilde x}^2}\Bigr)
\Bigl(\widetilde x-{\partial \widetilde\phi\over \partial \tilde x}
-{1\over \widetilde x}{\partial^2 \widetilde\phi\over \partial \theta^2}\Bigr)
-{1\over \widetilde x}\Bigl({1\over \widetilde x}
{\partial \widetilde\phi\over \partial \theta}
-{\partial^2 \widetilde\phi\over \partial \tilde x\partial \theta}
\Bigr)^2\biggr],
\label{Jacobzeroex}
\end{eqnarray}
where we used Eqs.~(\ref{explicita})$\sim$(\ref{explicite}). 
This is the exact approach (a) for the magnification. 

On the other hand, in the case of the perturbative approach (b),
we have
\begin{eqnarray}
&&J=\Bigl(1-{\partial^2 \widetilde\phi_0\over \partial \tilde x^2}\Bigr)
\times\Bigl(\widetilde {\delta x}
-\widetilde{\delta x}{\partial^2 \widetilde\phi_0\over \partial \tilde x^2}
-{\partial  \widetilde{\delta \phi} \over \partial \tilde x}
-{\partial^2 \widetilde{\delta\phi} \over \partial \theta^2}
\Bigr)\Bigr|_{\widetilde x=1},
\label{perturbativeJ}
\end{eqnarray}
where 
we used the condition of the Einstein ring.
$\partial\widetilde \phi_0/ \partial \tilde x|_{\widetilde x=1}=1$.
Note that 
$\widetilde\phi(\widetilde x,\theta)=\widetilde \phi_0(\widetilde x)+\widetilde{\delta\phi}(\widetilde x,\theta)$ and $\widetilde x=1+\widetilde{\delta x}$.
The formulas (\ref{dxxphi}) $\sim$(\ref{dttdeltaphi}) are used 
for the perturbative approach (b), 
and (\ref{dxdeltaphiap})$\sim$(\ref{appendoox})
for the approximate approach (c), respectively.

The upper right panels (a)$\sim$(d) of Fig.~\ref{figI} show
the magnification of the circumference of lensed image as a function 
of the angular coordinate of the image plane. 
The panels (a)$\sim$(c) correspond to the three approaches (a)$\sim$(c), 
respectively. The panel (d) plots the three approaches. 
The model parameters are the same as those of the upper left panels for 
the lensed image. 

\subsection{Critical line}
The definition of the critical curve is $J=0$ on the image plane.
In the exact approach (a), we solve 
\begin{eqnarray}
\widetilde x={\partial \widetilde\phi\over \partial \tilde x}
+{1\over \widetilde x}{\partial^2 \widetilde\phi\over \partial \theta^2}
+{1\over \widetilde x}\Bigl(1-{\partial^2 \widetilde\phi\over 
\partial {\tilde x}^2}\Bigr)^{-1}\Bigl({1\over \widetilde x}
{\partial \widetilde\phi\over \partial \theta}
-{\partial^2 \widetilde\phi\over \partial \tilde x\partial \theta}
\Bigr)^2,
\label{clex}
\end{eqnarray}
which is obtained using Eq.~(\ref{Jacobzeroex}).

In the perturbative approach (b), the critical line is 
\begin{eqnarray}
 \widetilde{\delta x}= {1\over 1- \partial^2\widetilde{\phi_0}/\partial \tilde x^2}\left(
  {\partial \widetilde{\delta\phi}\over \partial \tilde x}
 +{\partial^2 \widetilde{\delta\phi}\over \partial \theta^2}
\right)\biggr|_{\widetilde{x}=1},
\label{clap}
\end{eqnarray}
from $J=0$ with Eq.~(\ref{perturbativeJ}). This equation 
is the same as Eq.~(30) in the reference by Alard \cite{Alard07}.
In the perturbative approach (b), we use  
Eqs.~(\ref{dxdeltaphi}) and (\ref{dttdeltaphi}), 
while we use Eqs.~(\ref{dxdeltaphiap})
and (\ref{appendoox}) in the approximate approach (c).

The lower left panels (a)$\sim$(d) of Fig.~\ref{figI} show
the critical line (solid curve) on the image plane.
The panels (a)$\sim$(c) correspond to the three approaches (a)$\sim$(c), 
respectively, and the panel (d) plots the three approaches. 
The dotted circle is the Einstein ring.
The model parameters are the same as those of the other panels of
Fig.~\ref{figI}.

\subsection{Caustics}
The caustics are defined by $J=0$ on the source plane, which 
can be mapped from the critical line on the image plane by 
the lens equation. The caustics are important in understanding
the nature of deformation of an Einstein ring.

In the exact approach (a), the caustics are obtained by substituting
the solution of Eq.~(\ref{clex}) into (\ref{lenseqexx}) and
(\ref{lenseqexy}).

In the perturbative approach (b), 
the caustics are given by
\begin{eqnarray}
&&\widetilde {\delta y_1}={\partial^2 \widetilde{\delta \phi}\over \partial\theta^2}\bigg|_{\widetilde x=1}\cos\theta+
{\partial \widetilde{\delta \phi}\over \partial\theta}\bigg|_{\widetilde
x=1}\sin\theta,
\label{caua}
 \\ &&
\widetilde {\delta y_2}={\partial^2 \widetilde{\delta \phi}\over \partial\theta^2}\bigg|_{\widetilde x=1}\sin\theta-
{\partial \widetilde{\delta \phi}\over \partial\theta}
\bigg|_{\widetilde x=1}\cos\theta,
\label{caub}
\end{eqnarray}
which are obtained by substituting Eq.(\ref{clap})
into the lens equation (\ref{leneqptx}) and (\ref{leneqpty}). 
Eqs.~(\ref{caua}) and (\ref{caub}) are the same as 
Eq.~(31) in the reference by Alard\cite{Alard07}, 
where $x_s$ and $y_s$ are used instead of $\widetilde{\delta y_{1}}$ 
and $\widetilde{\delta y_{2}}$, respectively.
In the approximate approach (c), we use (\ref{dthetadeltaphiap}) 
and (\ref{appendoox}).

The lower right panels (a)$\sim$(d) of Fig.~\ref{figI} show
the caustics (solid curves) on the source plane. 
The panels (a)$\sim$(c) correspond to the three approaches  (a)$\sim$(c), 
respectively, 
and the panel (d) plots all three approaches.
The  spherical 
circles are the circumference of the source. 
The model parameters are the same as those of the other panels of 
Fig.~\ref{figI}. 

\subsection{Typical configurations}
\begin{figure}[hb]
\begin{center}
\begin{tabular}{cc}
\includegraphics[width=61mm, height=61mm]{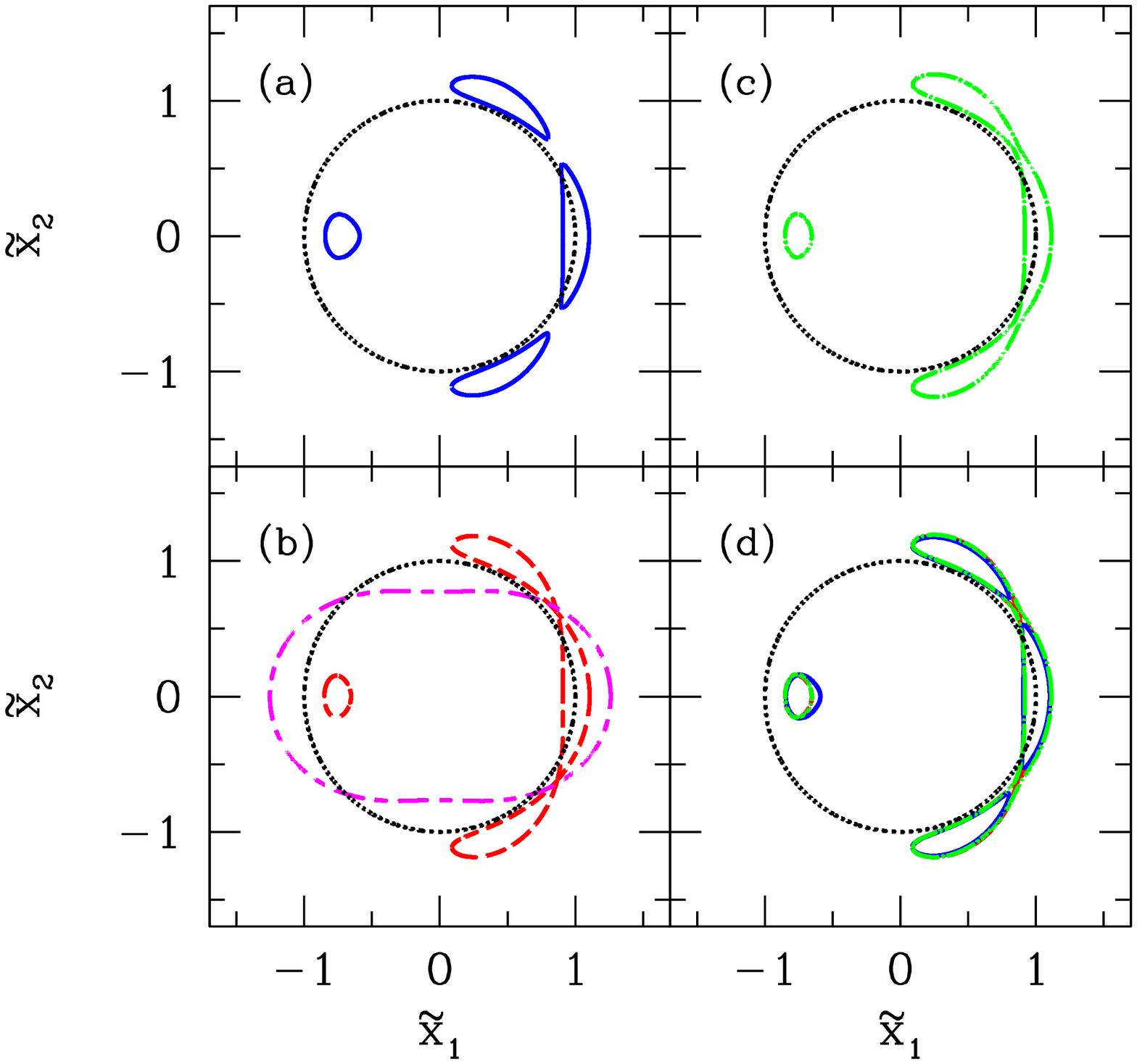}
&
\includegraphics[width=61mm, height=61mm]{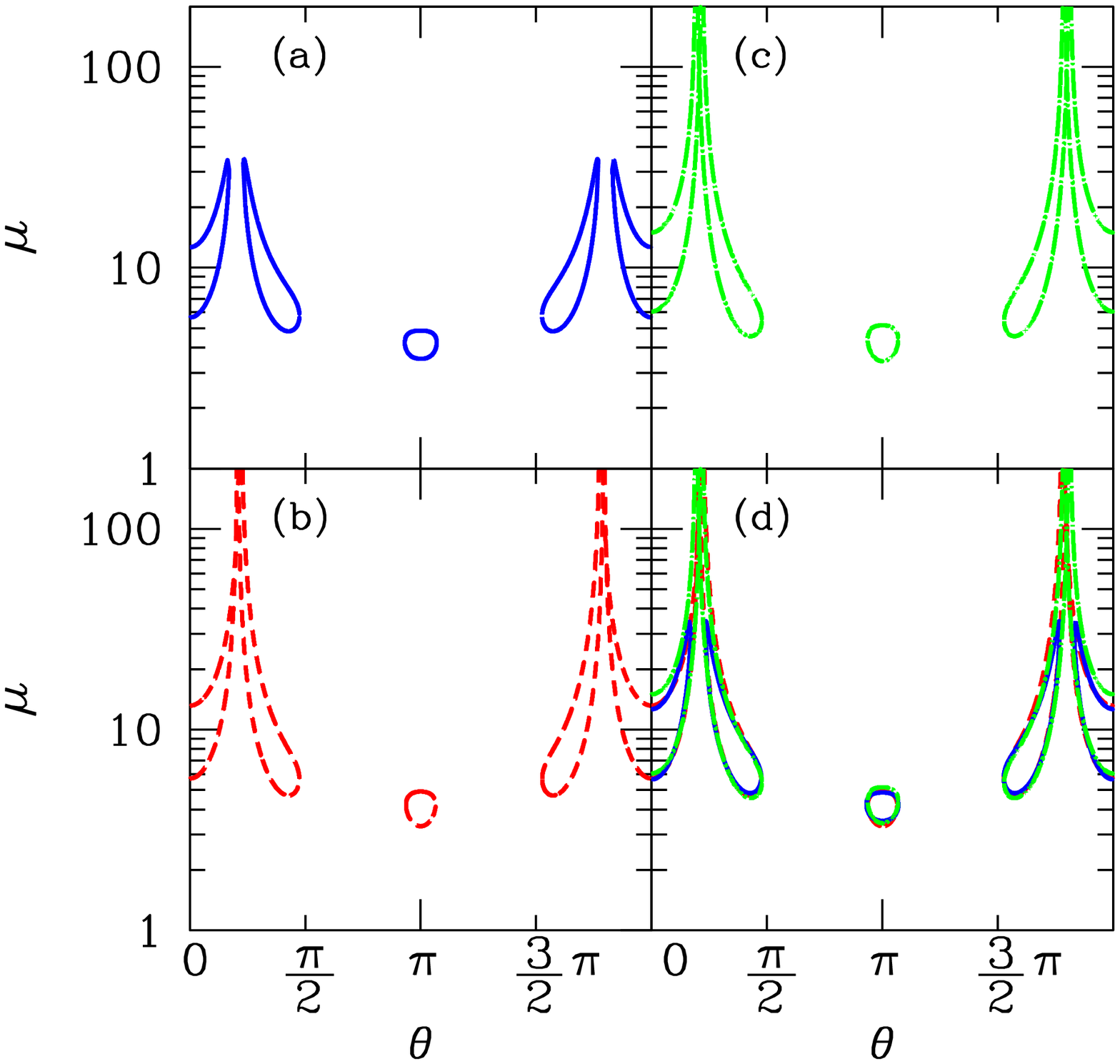}
\\
\includegraphics[width=61mm, height=61mm]{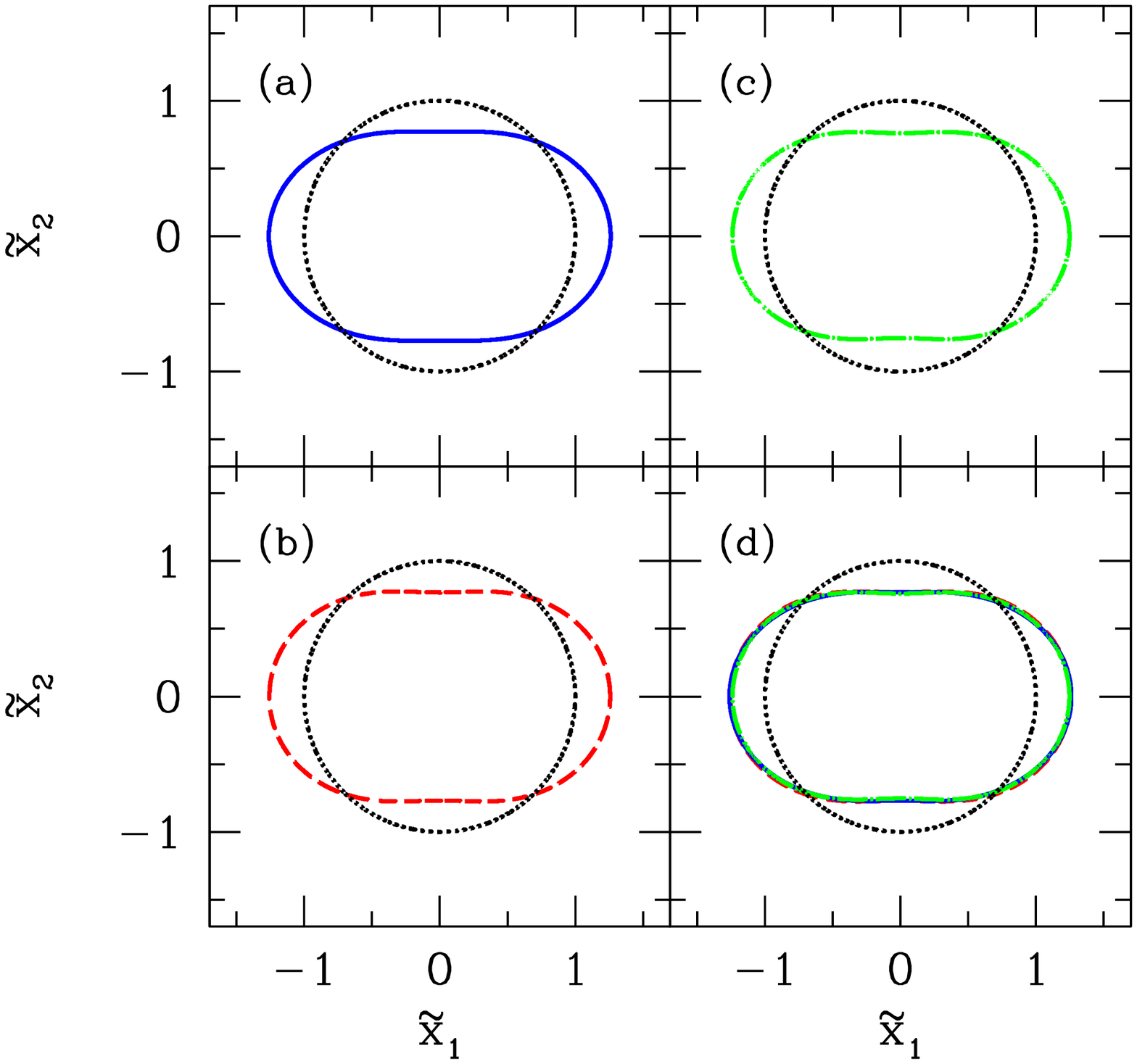}
&
\includegraphics[width=61mm, height=61mm]{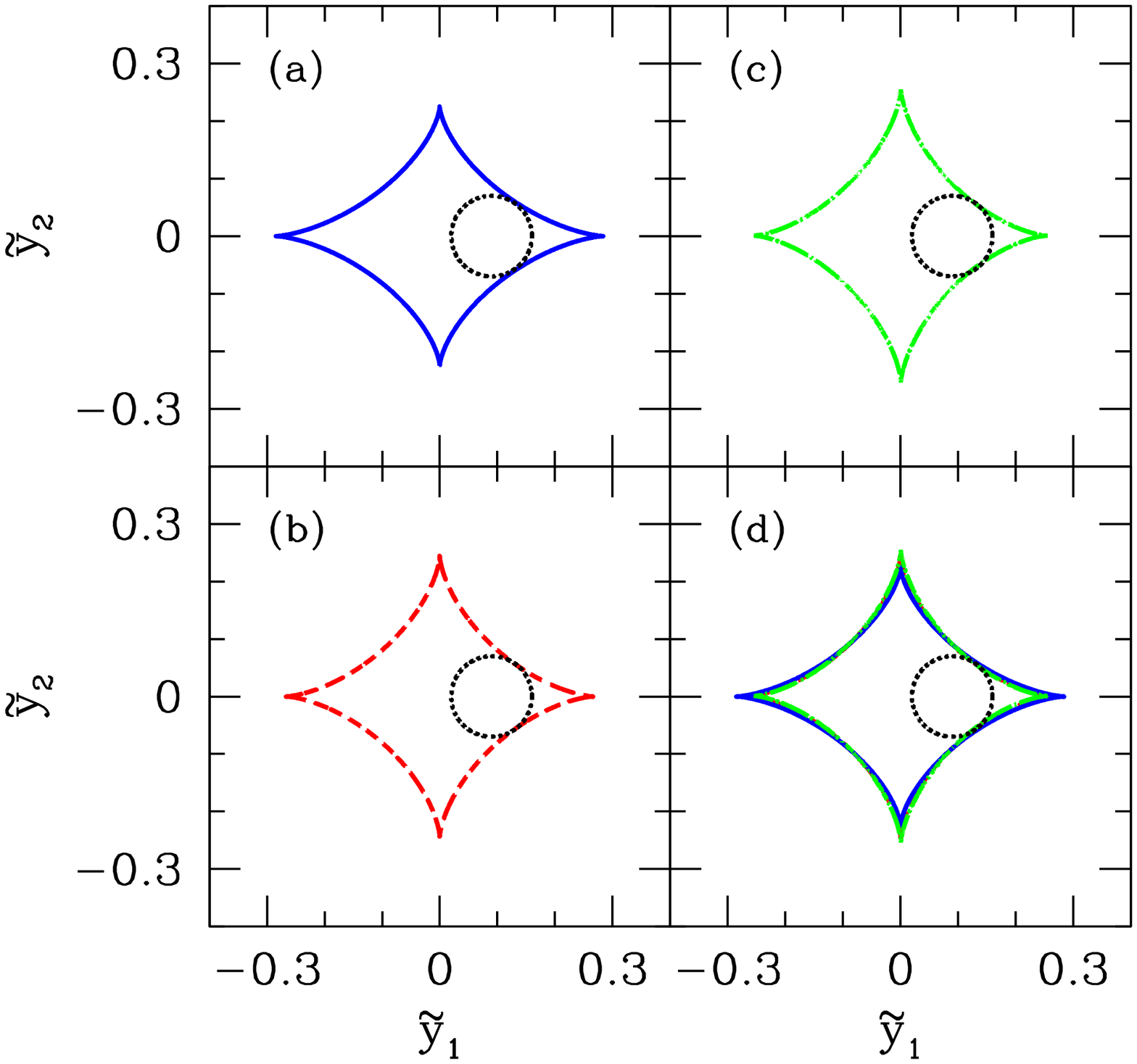}
\end{tabular}
\end{center}
\caption{Same as Fig.~\ref{figI} but with the parameters 
$\widetilde {\delta r_s}=0.07$, $\widetilde {\delta y_{10}}=0.09$,
$\widetilde {\delta y_{20}}=0$, $\eta=0.1274$ and $u_0=0.5$.
 This is type II.}
\label{figII}
\end{figure}

\begin{figure}[hptb]
\begin{center}
\begin{tabular}{cc}
\includegraphics[width=61mm, height=61mm]{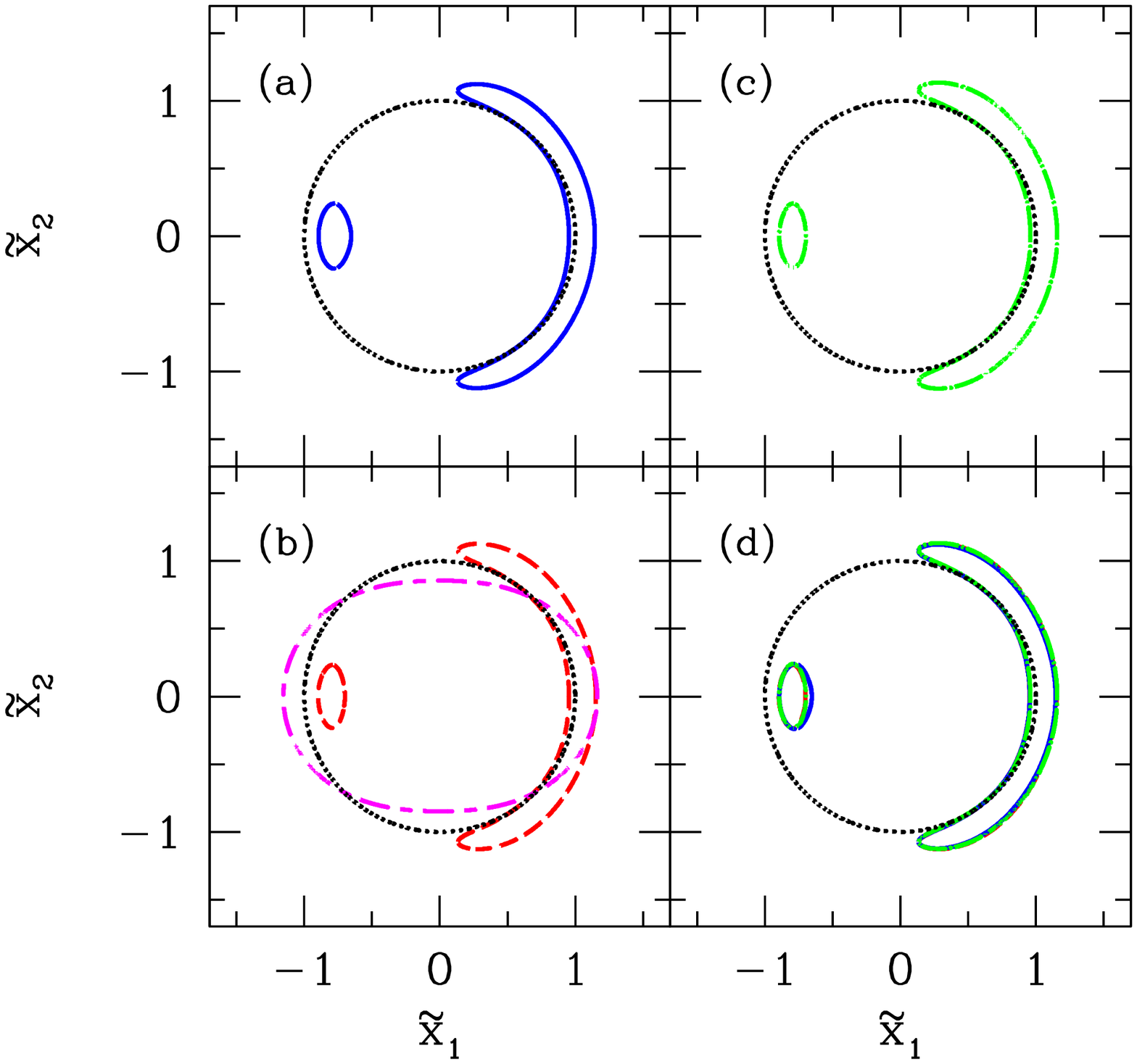}
&
\includegraphics[width=61mm, height=61mm]{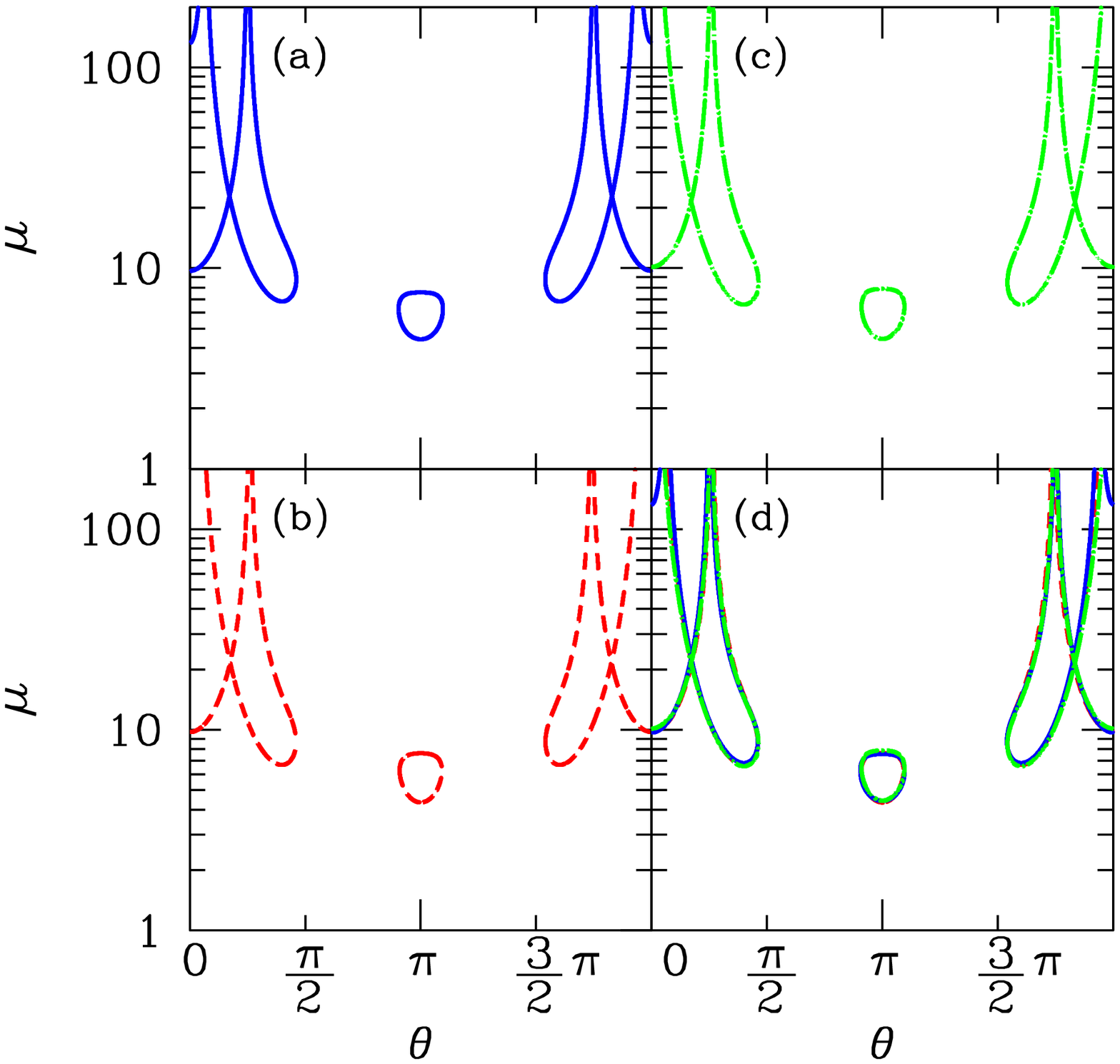}
\\
\includegraphics[width=61mm, height=61mm]{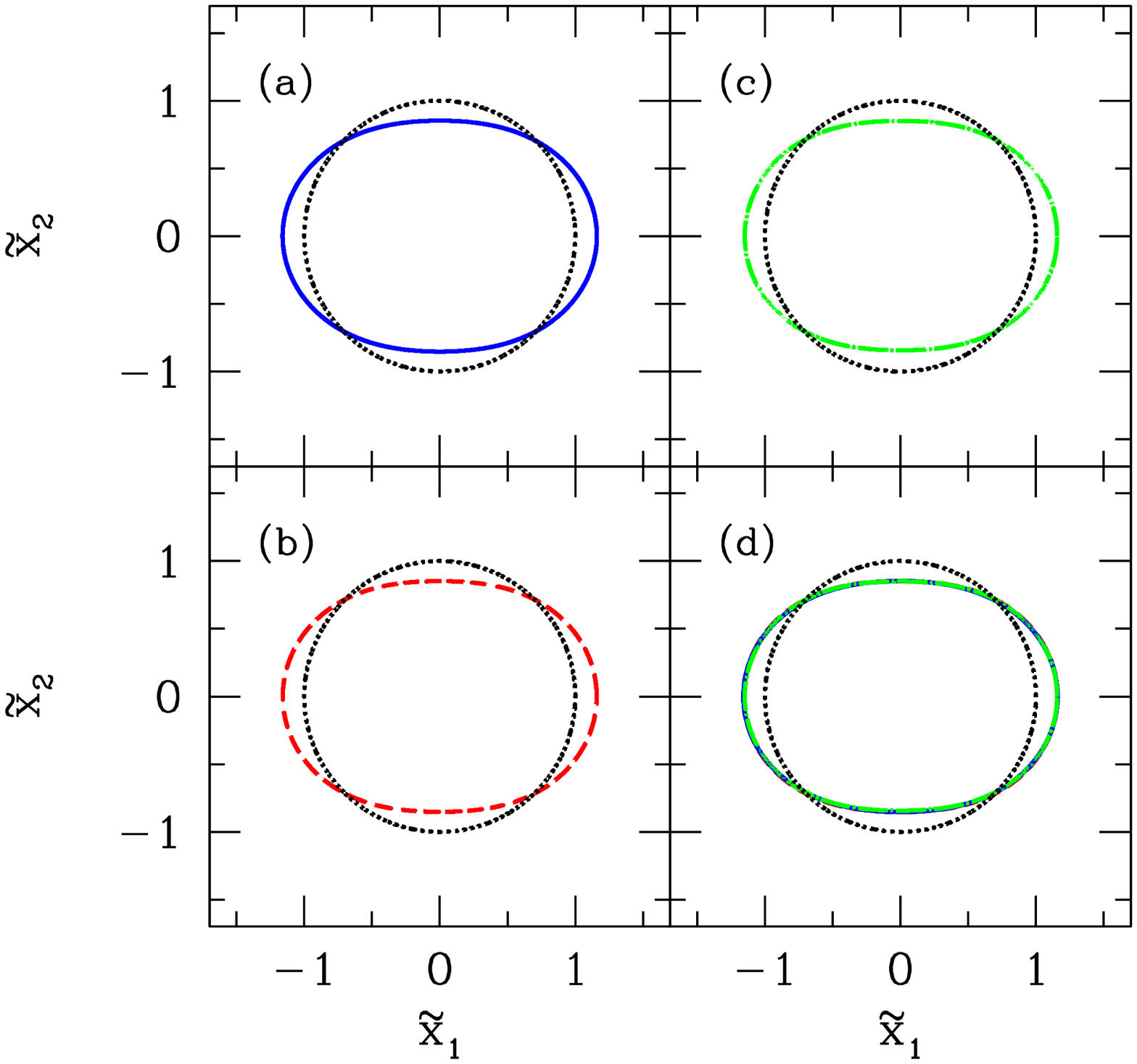}
&
\includegraphics[width=61mm, height=61mm]{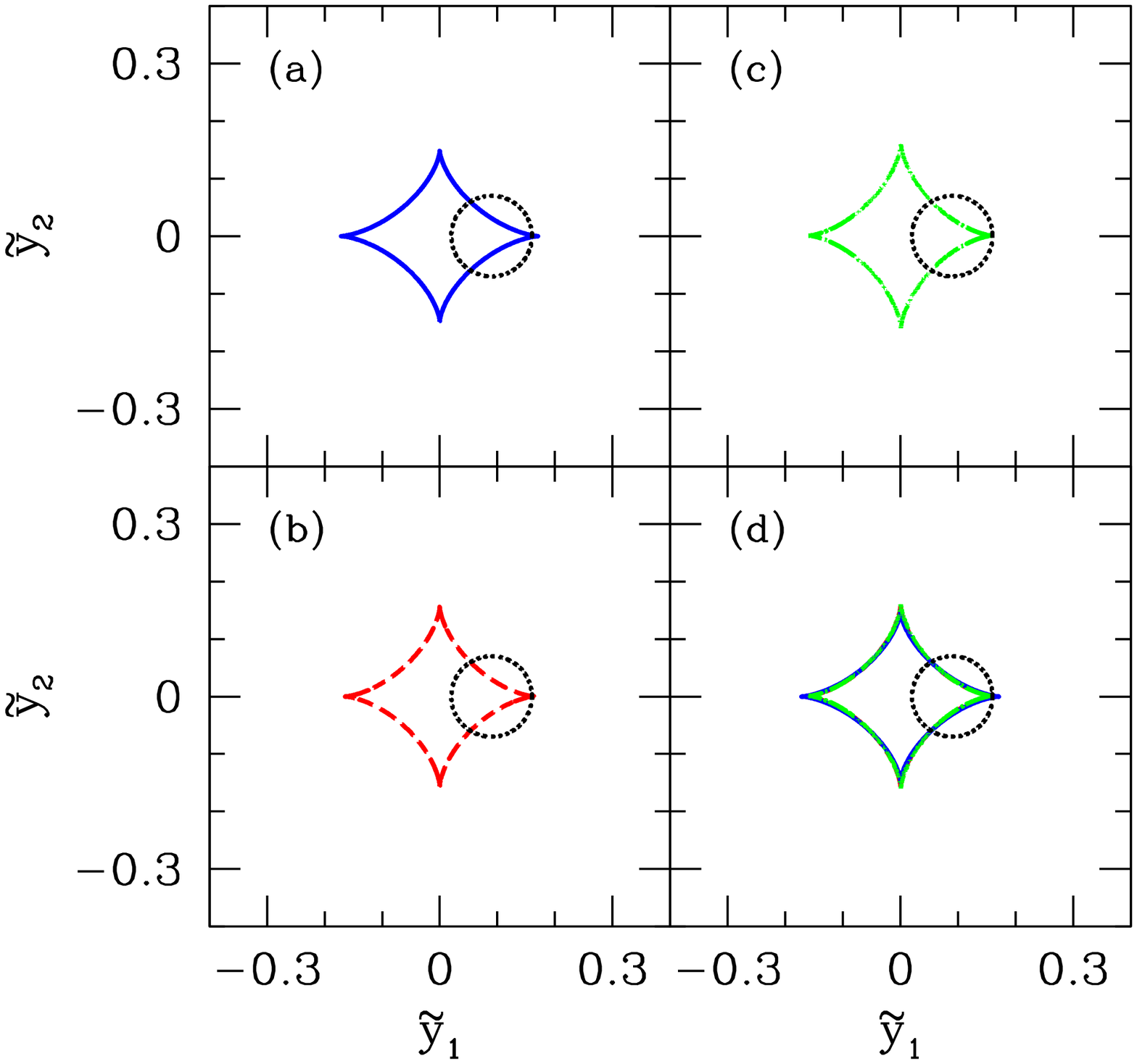}
\end{tabular}
\caption{Same as Fig.~\ref{figI} but with the parameters 
$\widetilde {\delta r_s}=0.07$, $\widetilde {\delta y_{10}}=0.09$,
$\widetilde {\delta y_{20}}=0$, $\eta=0.08$ and $u_0=0.5$.
 This is type III.}
\label{figIII}
\end{center}
\end{figure}

In Fig.~\ref{figI}, we adopted the parameters $\widetilde {\delta r_s}=0.07$, 
$\widetilde{\delta y_{10}}=0.09$, $\widetilde {\delta y_{20}}=0$, $\eta=0.15$ 
and $u_0=0.5$.
Figs.~\ref{figII} and \ref{figIII} are the same as 
Fig.~\ref{figI} but with $\eta=0.1274$ and $\eta=0.08$, 
respectively, instead of $\eta=0.15$.
The other parameters are the same as those of Fig.~\ref{figI}, 
then the source configuration of 
Figs.~\ref{figI},  \ref{figII}, and \ref{figIII} are the same.
As $\eta$ becomes smaller, the size of the caustics 
becomes smaller, and the shape of the critical line becomes 
more spherical. 
Also, as $\eta$ becomes smaller, the right-side 
three separated images become to merge.  
Fig.~\ref{figII} is the critical configuration when the merger 
occurs. 
One can observe that the merger occurs when the circumference of 
the source contacts with the caustics. 
In the upper right panels of Figs.~\ref{figII} 
and \ref{figIII}, the large enhancement of the 
magnification appears. 
This reflects the facts that the magnification diverges
when a source is on the caustics and when the image crosses the 
critical line. Figs.~\ref{figI}, \ref{figII} and \ref{figIII} 
represent typical types of lensed image, which we call type I, II 
and III, respectively.

\begin{figure}[hptb]
\begin{center}
\begin{tabular}{cc}
\includegraphics[width=61mm, height=61mm]{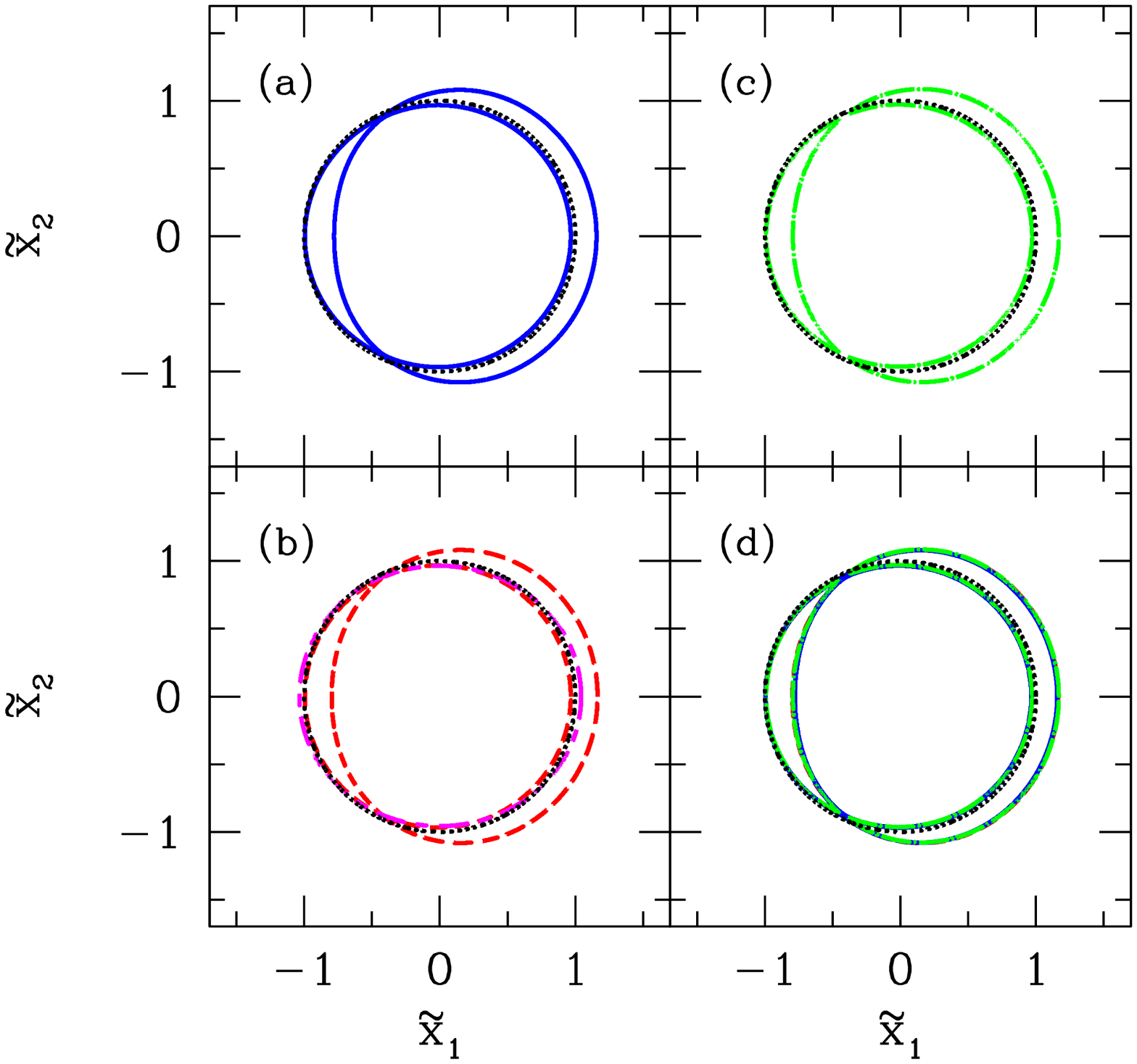}
&
\includegraphics[width=61mm, height=61mm]{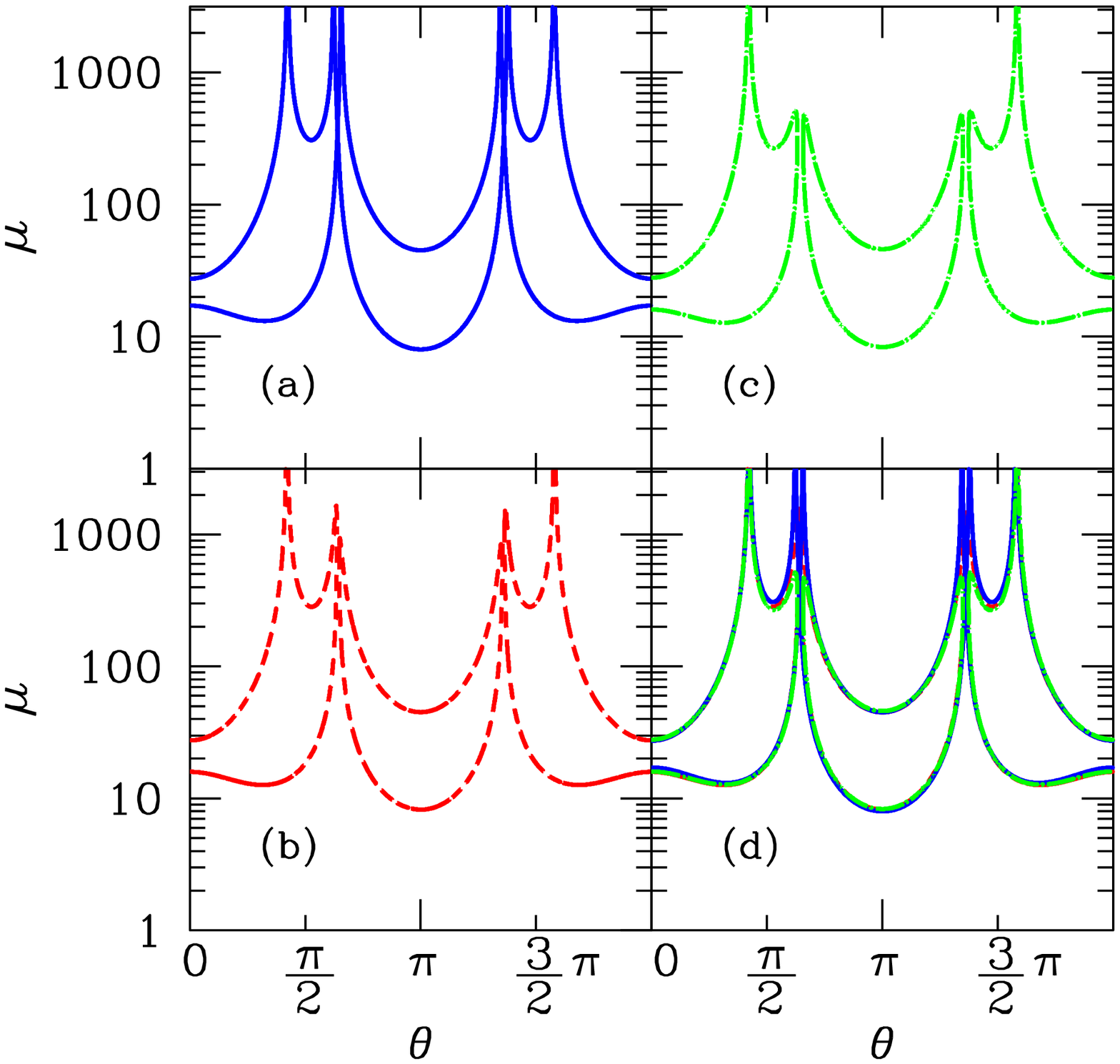}
\\
\includegraphics[width=61mm, height=61mm]{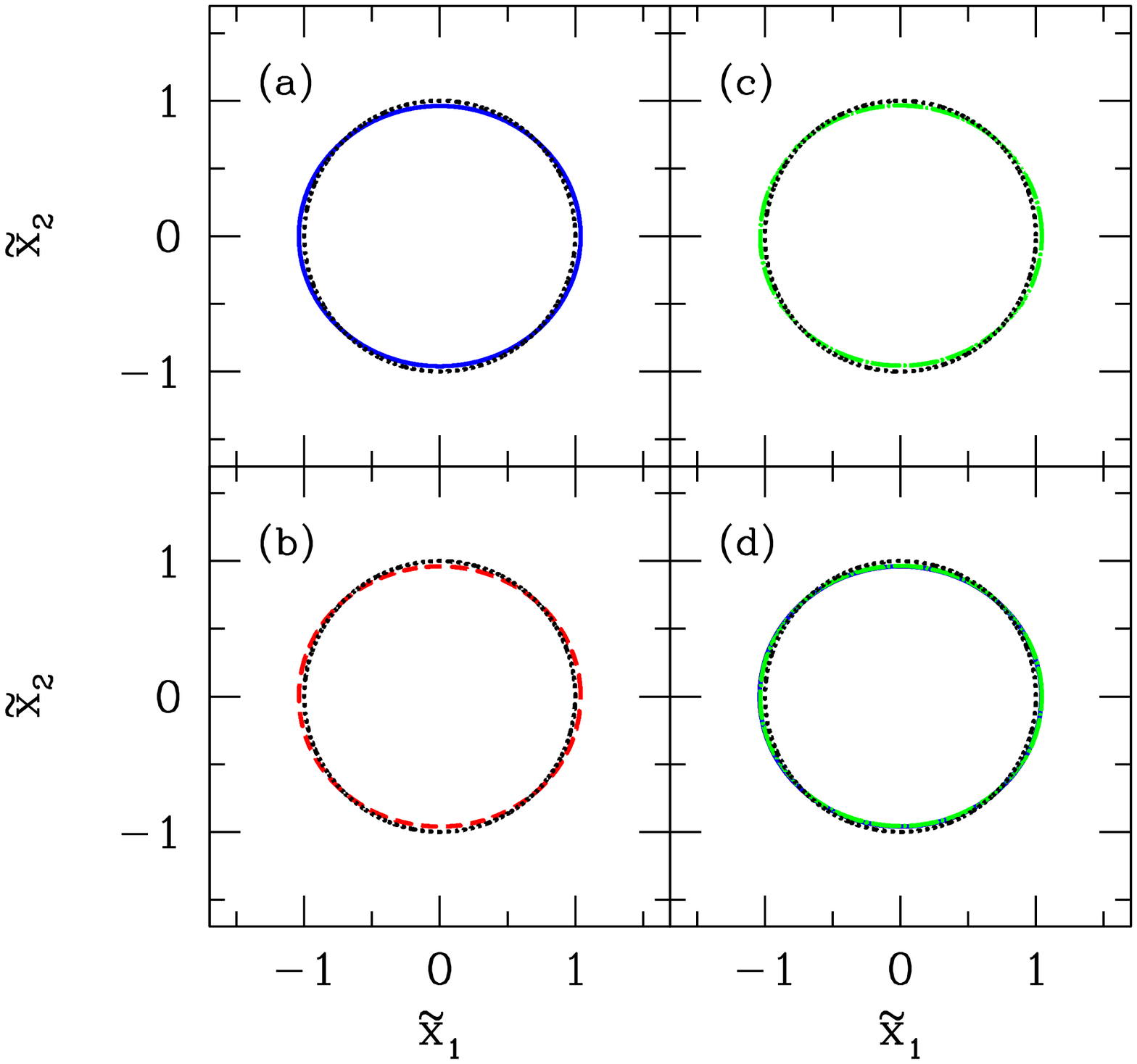}
&
\includegraphics[width=61mm, height=61mm]{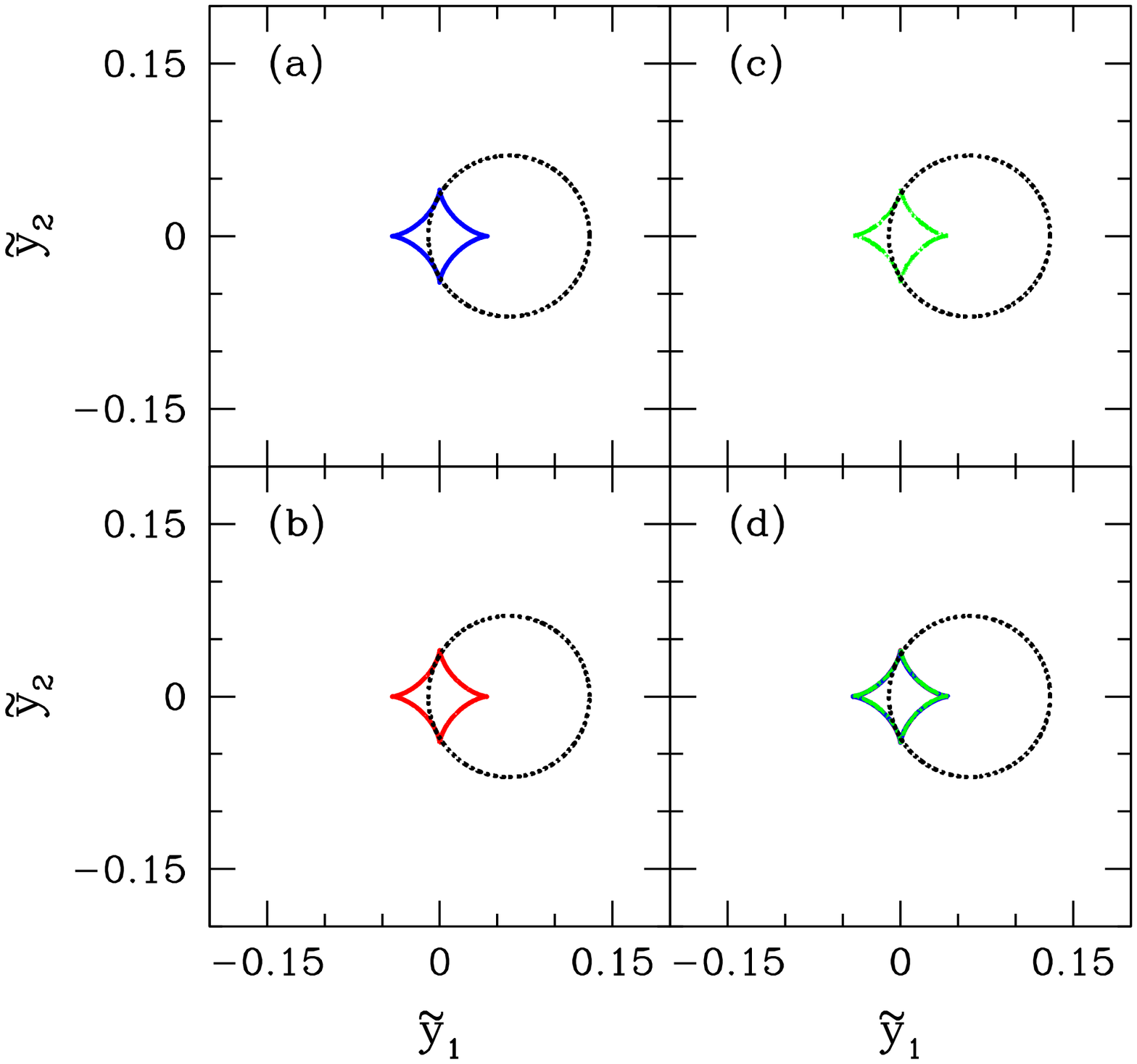}
\end{tabular}
\caption{Same as Fig.~\ref{figI} but with the parameters 
$\widetilde {\delta r_s}=0.07$, $\widetilde {\delta y_{10}}=0.06$,
$\widetilde {\delta y_{20}}=0$, $\eta=0.0205$ and $u_0=0.5$.
 This is type IV.
}
\label{figIV}
\end{center}
\end{figure}

\begin{figure}[hptb]
\begin{center}
\begin{tabular}{cc}
\includegraphics[width=61mm, height=61mm]{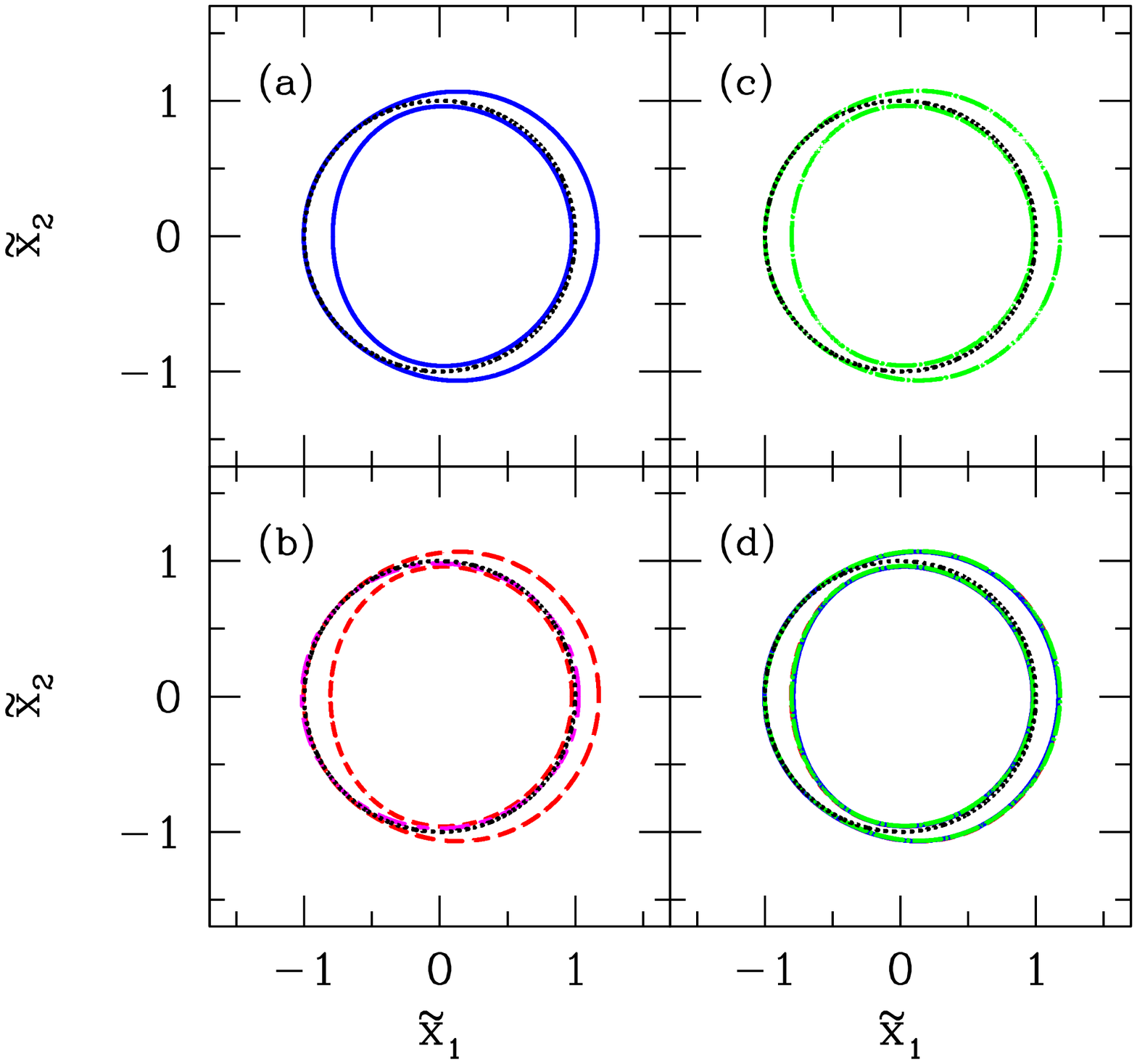}
&
\includegraphics[width=61mm, height=61mm]{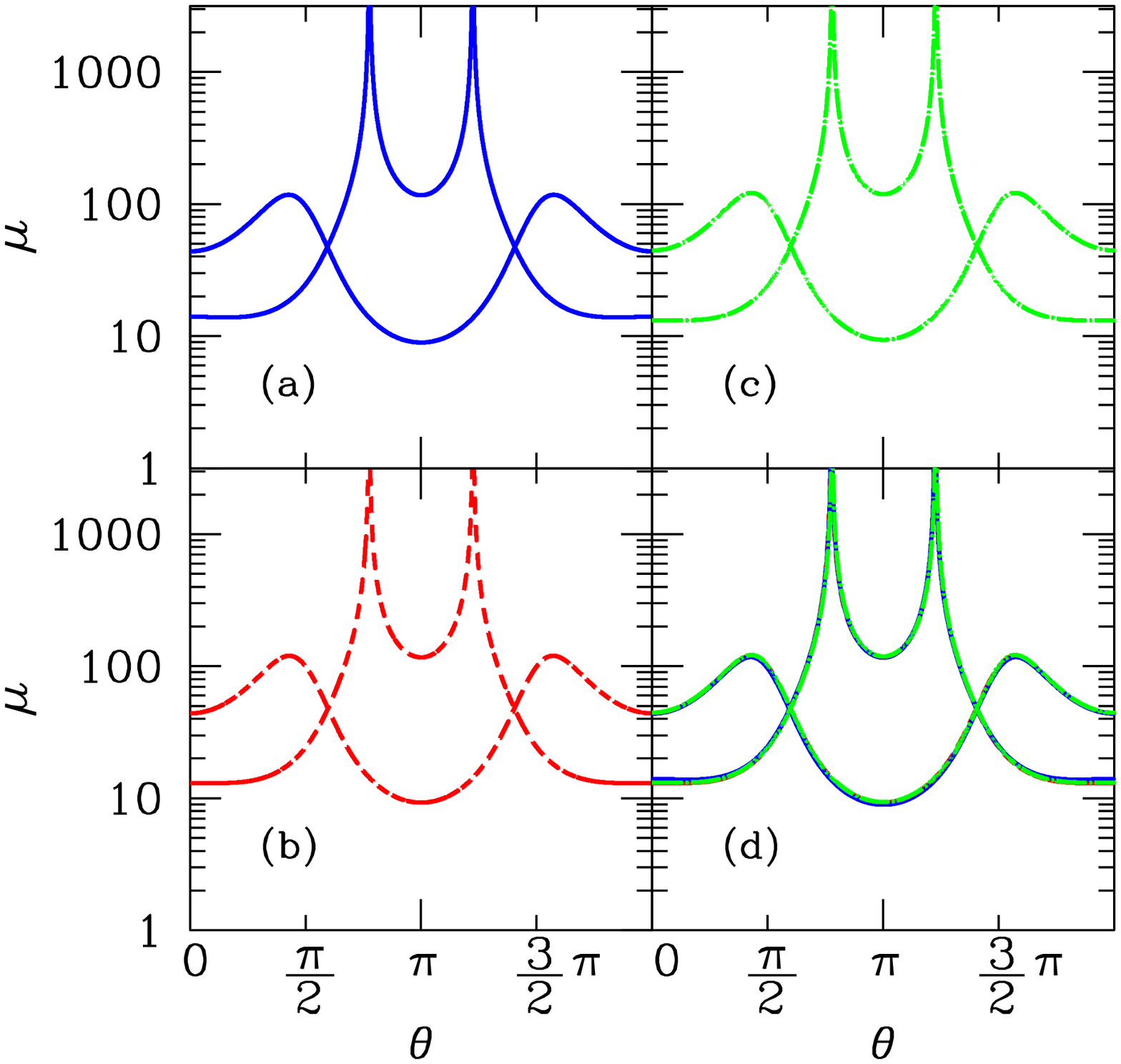}
\\
\includegraphics[width=61mm, height=61mm]{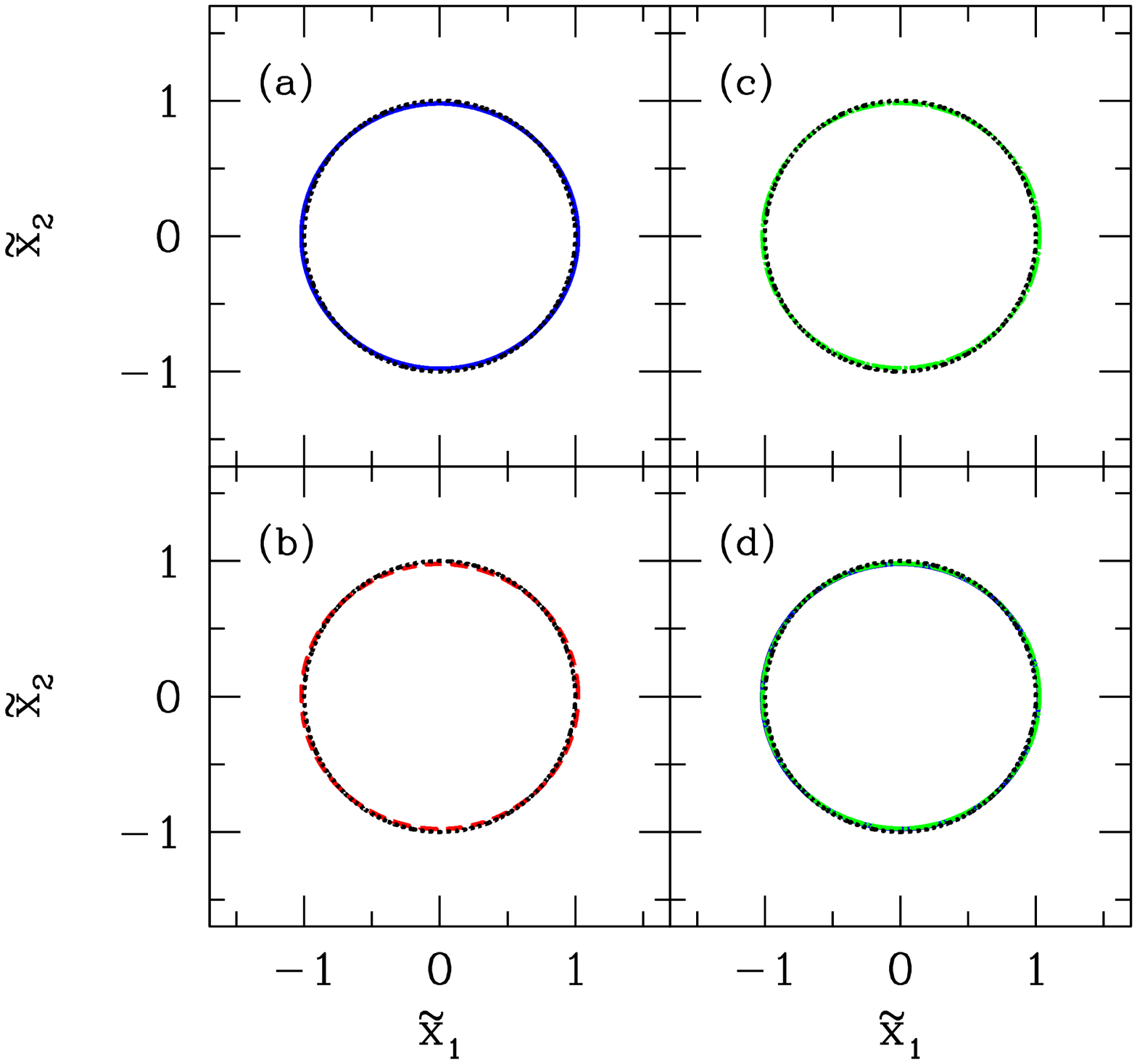}
&
\includegraphics[width=61mm, height=61mm]{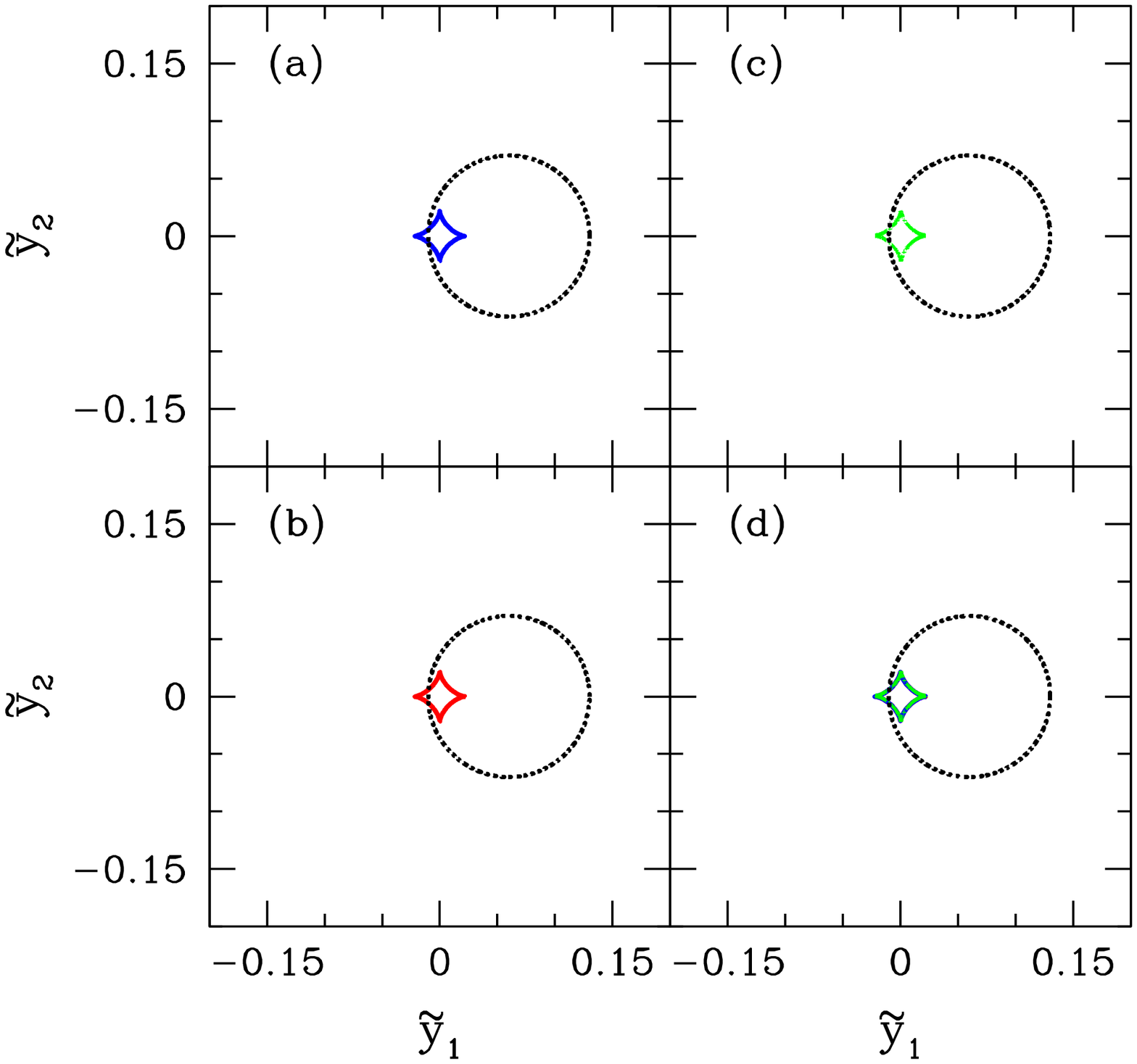}
\end{tabular}
\caption{Same as Fig.~\ref{figI} but with the parameters 
$\widetilde {\delta r_s}=0.07$, $\widetilde {\delta y_{10}}=0.06$,
$\widetilde {\delta y_{20}}=0$, $\eta=0.011$ and $u_0=0.5$.
 This is type V. }
\label{figV}
\end{center}
\end{figure}

As $\eta$ becomes smaller furthermore, the left-side image and the 
right-side image become elongated, and form a ring, as is demonstrated
in Figs.~\ref{figIV} and \ref{figV}.
The parameters of Fig.~\ref{figIV} are $\widetilde {\delta r_s}=0.07$, 
$\widetilde{\delta y_{10}}=0.06$, $\widetilde {\delta y_{20}}=0$, $\eta=0.0205$ 
and $u_0=0.5$. Fig.~\ref{figV} is the same as Fig.~\ref{figIV} but 
with the different value of $\eta=0.011$, instead of $\eta=0.0205$. 
These two images are almost rings, i.e., Einstein rings with finite 
width. 
Fig.~\ref{figIV} is the critical configuration when a ring 
is formed. 
Figs.~\ref{figIV} and \ref{figV} 
represent typical types of lensed image of a ring, 
which we call type IV and V, respectively.
In these cases, the size of the caustics is smaller than the source size, 
which is clearly shown in Figs.~\ref{figIV} and \ref{figV}. 
We note that the critical configuration, type IV, appears
when the circumference of the source comes in contact with 
the caustics at the left-side caustic. 
The divergence of the magnification appears
when the source overlaps the caustics 
and when the image crosses the critical line.
 


\subsection{Validity of the perturbative approximation}
\begin{figure}[htb]
\begin{center}
\includegraphics[width=140mm, height=140mm]{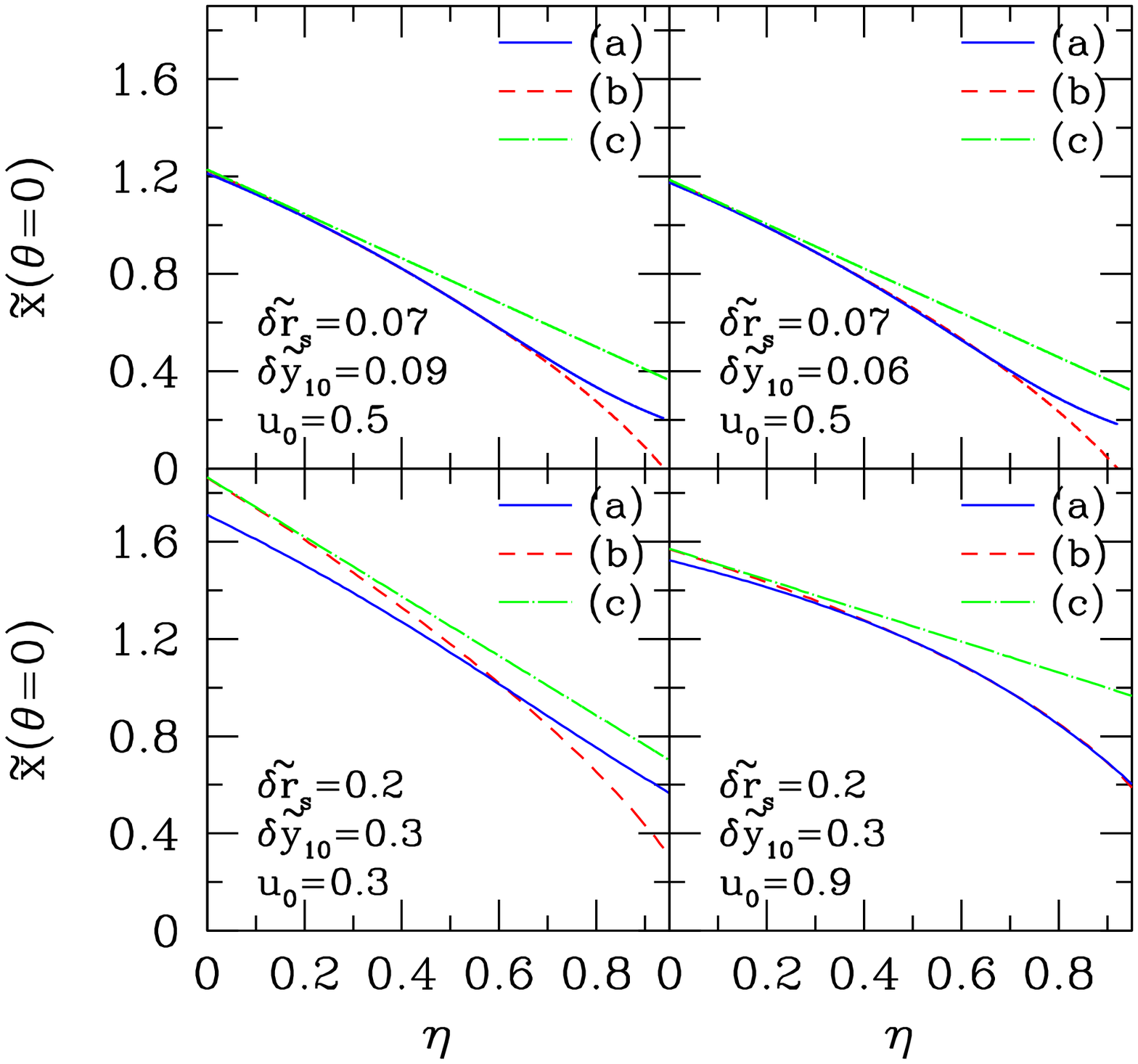}
\caption{Comparison of the solution of the lens equation
$\widetilde x(\theta=0)$ as a function of $\eta$. 
The solid curve is the exact approach (a), 
the dashed curve is the perturbative approach (b), 
and the long dash-dotted  curve is the approximate approach (c). 
The upper left panel adopted $\widetilde {\delta r_s}=0.07$, 
$\widetilde {\delta y_{10}}=0.09$,
$\widetilde {\delta y_{20}}=0$ and $u_0=0.5$.
The upper right panel is the same as the upper left panel but 
with the different value of 
$\widetilde {\delta y_{10}}=0.09$. 
The lower left (right) is  $\widetilde {\delta r_s}=0.2$ 
$\widetilde {\delta y_{10}}=0.3$, $u_0=0.3$ ($u_0=0.9$).}
\label{fig:xtheta0eta}
\end{center}
\end{figure}

\begin{figure}[htbp]
\begin{center}
\includegraphics[width=100mm, height=100mm]{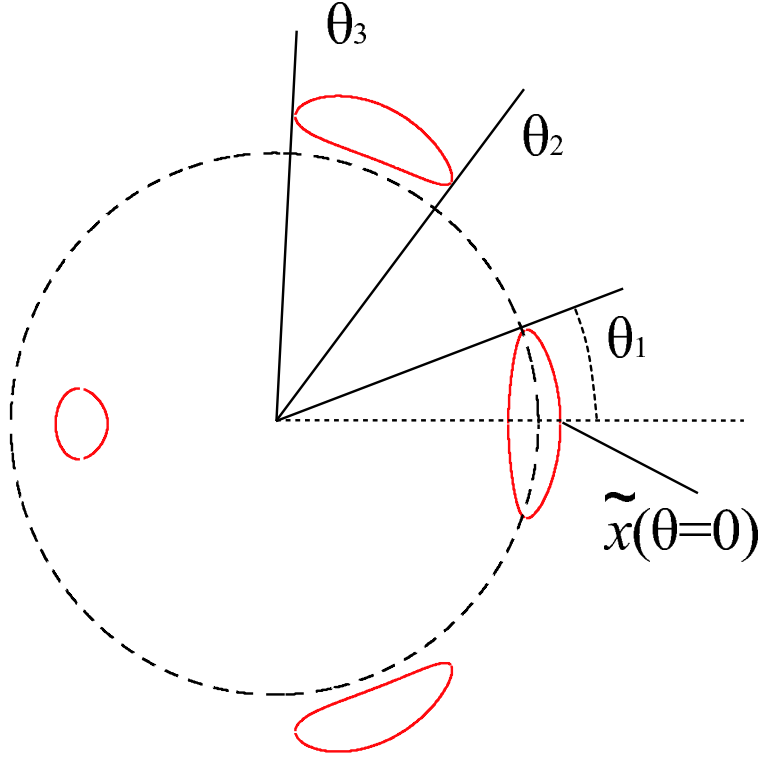}
\caption{Definition of the angles 
$\theta_1$, $\theta_2$,  and $\theta_3$, for the image of type I.}
\label{fig:imageexplain}
\end{center}
\end{figure}

Let us discuss about the validity of the perturbative approach
comparing with the exact approach. Fig.~\ref{fig:xtheta0eta} 
plots the outer position of the image at $\theta=0$, i.e., 
$\widetilde x(\theta=0)$, as a function of $\eta$.
(See also Fig.~\ref{fig:imageexplain}.)
The curves labelled by (a), (b) and (c) correspond to the three approaches.
In the exact approach (a), $\widetilde x(\theta=0)$ is given by solving 
Eq.~(\ref{exactimage}).
In the perturbative approach (b) and the approximate approach (c), 
$\widetilde x(\theta=0)=1+\widetilde{\delta x}(\theta=0)$, 
where $\widetilde{\delta x}(\theta=0)$
is given from Eq.~(\ref{dreqap}) with the formulas in the appendix B
and in the appendix C, respectively.
We are considering the circumference of a circular source, and 
$\widetilde x(\theta=0)$ has the two solutions, which correspond
to the two solutions of the sign $\pm$ in Eq.~(\ref{exactimage})
in the exact approach (a) and Eq.~(\ref{dreqap}) 
in the approaches (b) and (c). 
Fig.~\ref{fig:xtheta0eta} plots the outer solution, which corresponds
to the solution of the sign $+$ in Eqs.~(\ref{exactimage}) and (\ref{dreqap}). 
In the limit, $\eta \rightarrow0$, (b) is equivalent to (c). 
The difference between (a) and (b) (or (c)) in the limit $\eta 
\rightarrow0$, comes from a limitation of the perturbative method.
As one can see from Fig.~\ref{fig:xtheta0eta}, the difference 
between (a) and (b) becomes large for $\eta\simgt0.5$, where
the perturbative approach breaks down.
The similar result was obtained in Figure 1 
in reference by Alard (2007) \cite{Alard07}. 
Besides comparing $\widetilde x(\theta=0)$, it is useful 
to compare the position of the image edge as examined 
in Fig.~8 in the below
(See also the reference by Peirani, et~al.(2008)\cite{Peirani08}).

\begin{figure}[phtb]
\begin{center}
\includegraphics[width=120mm, height=120mm]{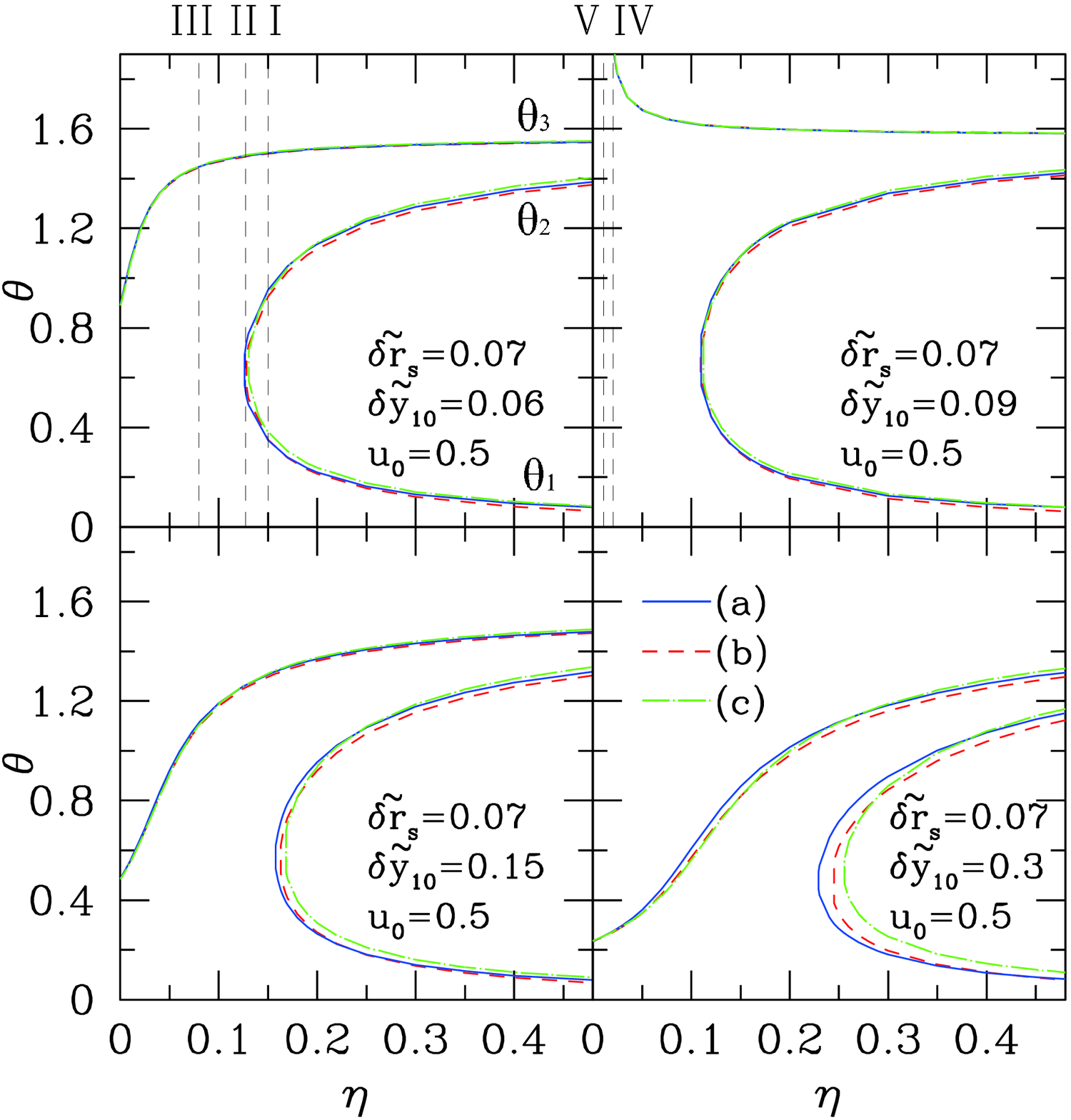}
\end{center}
\caption{ $\theta_1$, $\theta_2$ and $\theta_3$ as a function of $\eta$.
In each panel, the top curve is $\theta_3$. 
As $\eta$ becomes small, $\theta_1$ and $\theta_2$ merge,
when the critical configuration (type II) appears.
The solid curve is the exact approach (a), 
the dashed curve is the perturbative approach (b), 
and the long dash-dotted curve is the approximate approach (c).
The dashed vertical lines are the values of $\eta$ adopted 
in Figs.~\ref{figI}$\sim$\ref{figV}, as labelled in this figure.  
In this figure we adopt  $\widetilde{\delta r_s}=0.07$, 
$\widetilde{\delta y_{20}}=0$ and $u_0=0.5$, and 
$\widetilde{\delta y_{10}}=0.09$ (upper left panel);
$\widetilde{\delta y_{10}}=0.06$ (upper right panel);
$\widetilde{\delta y_{10}}=0.15$ (lower left panel);
$\widetilde{\delta y_{10}}=0.3$ (lower right panel).}
\label{figthetaeta}
\end{figure}

Next, let us focus on the critical configurations 
where the number of the images changes. 
Four images of the type I change to two images of the type III
as $\eta$ becomes smaller. The type II is the critical configuration. 
The two images of the type III changes to a ring configuration 
of the type V as $\eta$ becomes smaller. 
The type IV is also the critical configuration.
Fig.~\ref{figthetaeta} examines these critical behaviours. 
Fig.~\ref{figthetaeta} shows the
angles of the edge of the image, $\theta_1$, $\theta_2$ and $\theta_3$, 
which are defined using Fig.~\ref{fig:imageexplain}, as a function 
of $\eta$.
In each panel, we plot $\theta_1$, $\theta_2$ and $\theta_3$,
and the top curve is $\theta_3$.
As $\eta$ becomes small, $\theta_2$ and $\theta_1$ merge, when
the critical configuration (type II) appears. 
We refer $\eta_{\rm II}$ as the 
critical value of $\eta$ when $\theta_2$ and $\theta_1$ merge. 
For $\eta<\eta_{\rm II}$, we have no solution for $\theta_1$ and 
$\theta_2$, 
In the upper left panel of Fig.~\ref{figthetaeta}, the 
vertical dashed line labelled by I, II and III, are the value 
of $\eta$, adopted in Fig.~\ref{figI},~\ref{figII} and ~\ref{figIII},
respectively. 
The solid curve is the exact approach (a), 
the dashed curve is the perturbative approach (b), 
and the long dash-dotted curve is the approximate approach (c). 
Note that the three approaches (a), (b) and (c) agree very well.

The upper right panel of Fig.~\ref{figthetaeta} is the same as
the upper left panel, but with 
$\widetilde {\delta y_{10}}=0.06$, instead of $\widetilde {\delta y_{10}}=0.09$.
Similarly, the lower left (right) panel is the same as the upper left panel, 
but with $\widetilde {\delta y_{10}}=0.15$  ($\widetilde {\delta y_{10}}=0.3$). 
Thus, the agreement between the three approaches is better for  
smaller values of $\widetilde {\delta y_{10}}$.
Note also that the critical value $\eta_{\rm II}$ becomes larger 
as $\widetilde {\delta y_{10}}$ becomes larger.

In the upper right panel of Fig.~\ref{figthetaeta}, the 
vertical dashed line labelled by IV and V, is the value
of $\eta$, adopted in Figs.~\ref{figIV} and ~\ref{figV},
respectively.
In this panel, $\theta_3$ increases as $\eta$ becomes small.
We define $\eta_{\rm IV}$ by the smallest value of $\eta$ with
which we can find the solution of $\theta_3$.
The type IV critical configuration appears at $\eta=\eta_{\rm IV}$.
as shown by the vertical dashed line labelled by IV.  
%
The critical configuration of type IV appears only in
the upper right panel of Fig.~\ref{figthetaeta}, where a ring
image appears for $\eta<\eta_{\rm IV}$. 


\newpage

\newpage

\newpage
\section{Discussions}
As is demonstrated in the above, the approximate approach (c) is not 
so bad. 
An advantage
of the approximate approach (c) is that  the simplicity enables us to
investigate the lensing phenomena in an analytic way, as we will show in
this section. 

\subsection{Condition of the critical configurations}
First, we consider the condition that the critical configuration of 
type II appears, which is the transition point that the number of the 
images changes from four to two. 

\begin{figure}[htbp]
\includegraphics[width=80mm, height=100mm]{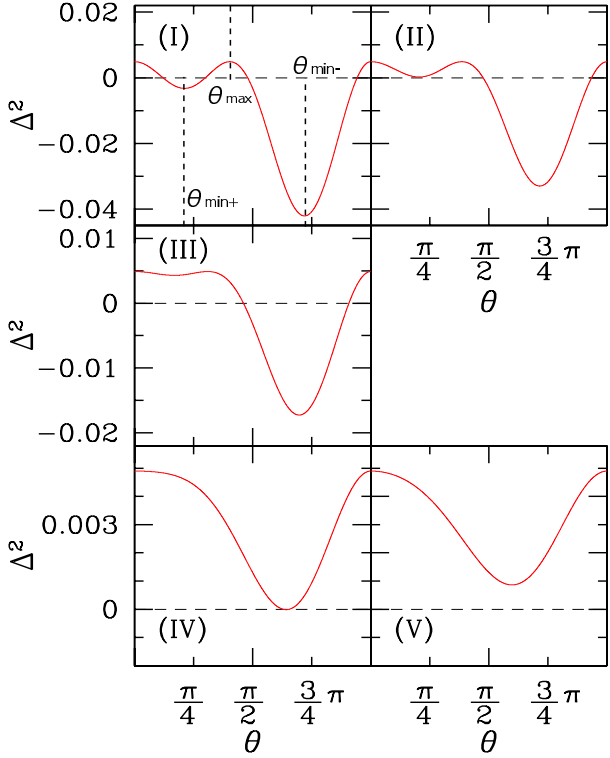}
\caption{${\widetilde\Delta}^2(\widetilde x=1,\theta)$
as a function of $\theta$ (see Eq.~(\ref{Discussa})). 
Here the parameters of each panel (I) $\sim$ (V) are the same 
as those of Figs.~\ref{figI} $\sim$ ~\ref{figV}, respectively.}
\label{figDiscussa}
\end{figure}

In the perturbative approach, the condition $\Delta^2(x_E,\theta)\geq0$
must be satisfied for the existence of the solution of the lens 
equation (see Eqs.~(\ref{dreq}) and (\ref{Delta2})).  
In the approximate approach (c), from Eqs.~(\ref{Delta2ap}) and 
(\ref{dthetadeltaphiap}), we have
\begin{eqnarray}
{\widetilde\Delta}^2(\widetilde x=1,\theta)=\widetilde{\delta r_s}^2-(\eta\sin2\theta
-\widetilde{\delta y_{10}}\sin\theta)^2,
\label{Discussa}
\end{eqnarray}
where we used $\widetilde{\delta y_{20}}=0$. Fig.~\ref{figDiscussa}
plots ${\widetilde\Delta}^2(\widetilde x=1,\theta)$ as a function
of $\theta$ for the five typical cases, I $\sim$ V, 
corresponding to Figs.~\ref{figI} $\sim$ \ref{figV}.
Note that our models satisfy ${\widetilde\Delta}^2(\widetilde x=1,\theta)
={\widetilde\Delta}^2(\widetilde x=1,2\pi-\theta)$.
Fig.~\ref{figDiscussa} only plots the range of $0\leq \theta \leq \pi$.
Lensed image appears under the condition 
${\widetilde\Delta}^2(\widetilde x=1,\theta)\geq0$.
The number of separated regions with 
${\widetilde\Delta}^2(\widetilde x=1,\theta)\geq0$ determines the 
number of the images. Also the zero points of 
${\widetilde\Delta}^2(\widetilde x=1,\theta)=0$ in the panel (I) of 
Fig.~\ref{figDiscussa} correspond to $\theta_1$, $\theta_2$ and $\theta_3$ 
in Fig.~\ref{fig:imageexplain}.

Let us first explain the local minimum and maximum points
of ${\widetilde\Delta}^2(\widetilde x=1,\theta)$, 
$\theta_{\rm min\pm}$ and $\theta_{\rm max}$, respectively,
(see the upper left panel of Fig.~\ref{figDiscussa}).
From (\ref{Discussa}), we have, 
\begin{eqnarray}
{d \over d\theta }{\widetilde\Delta}^2(\widetilde x=1,\theta)&=&
\sin\theta[-16\eta^2\cos^3\theta+12\eta\widetilde{\delta y_{10}}\cos^2\theta
\nonumber
\\&&+(8\eta^2-2\widetilde{\delta y_{10}}^2)\cos\theta-4\eta\widetilde{\delta y_{10}}]
\nonumber
\\
&=&-16\eta^2\sin\theta\left(\cos\theta-{\widetilde{\delta y_{10}}\over 2\eta}\right)
\nonumber
\\
&&\times
\left(\cos\theta-{\widetilde{\delta y_{10}}+\sqrt{32\eta^2+\widetilde{\delta y_{10}}^2}\over 8\eta}
\right)
\nonumber
\\&&\times\left(\cos\theta-{\widetilde{\delta y_{10}}-\sqrt{32\eta^2+\widetilde{\delta y_{10}}^2}\over 8\eta}
\right).
\end{eqnarray}
We find that ${\widetilde\Delta}^2(\widetilde x=1,\theta)$ has a local 
maximum in the range of $0<\theta<\pi$ at 
\begin{eqnarray}
\cos\theta_{\rm max}={\widetilde{\delta y_{10}}\over 2\eta}{},
\label{Discussd}
\end{eqnarray}
and the two local minimum at
\begin{eqnarray}
\cos\theta_{\rm min\pm}={1\over 8\eta} \left(
\widetilde{\delta y_{10}}\pm \sqrt{32\eta^2+\widetilde{\delta y_{10}}^2}
\right).
\label{Discussb}
\end{eqnarray}
The values of the local maximum and minimum are
\begin{eqnarray}
{\widetilde\Delta}^2(\widetilde x=1,\theta=\theta_{\rm max})={\widetilde{\delta r_s}}^2
\label{Discusse}
\end{eqnarray}
and 
\begin{eqnarray}
&&{\widetilde\Delta}^2(\widetilde x=1,\theta=\theta_{\rm min \pm})
\nonumber
\\
&&~~={1\over 128\eta^2}\bigg[
{\widetilde{\delta y_{10}}}^4-80{\widetilde{\delta y_{10}}}^2\eta^2-128\eta^4
\pm{\widetilde{\delta y_{10}}}^3\sqrt{32\eta^2+{\widetilde{\delta y_{10}}}^2}
\nonumber
\\&&~~\pm32\eta^2{\widetilde{\delta y_{10}}}\sqrt{32\eta^2+{\widetilde{\delta y_{10}}}^2}
+128{\widetilde{\delta r_s}}^2\eta^2
\bigg]\equiv{\widetilde\Delta}^2_{\pm},
\label{Discussc}
\end{eqnarray}
respectively.

Furthermore, ${\widetilde\Delta}^2(\widetilde x=1,\theta)$ also has 
other local maximum at $\theta=0$ and $\theta=\pi$. From
(\ref{Discussa}), we have
\begin{eqnarray}
{\widetilde\Delta}^2(\widetilde x=1,\theta=0)={\widetilde\Delta}^2(\widetilde x=1,\theta=\pi)={\widetilde{\delta r_s}}^2.
\label{Discussf}
\end{eqnarray}
This means that the thickness of the arcs is approximately determined by
$\widetilde{\delta r_s}$, which can be seen in 
Figs.~\ref{figI}$\sim$~\ref{figV}.

\begin{figure}[htbp]
\begin{center}
\includegraphics[width=70mm, height=70mm]{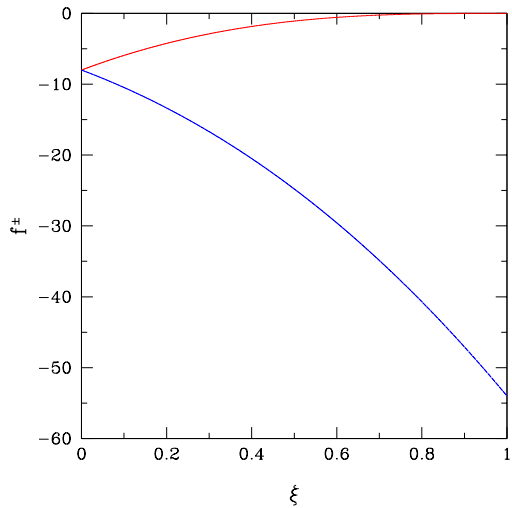}
\caption{$f^+$ (upper curve) and $f^-$ (lower curve) as a function of $\xi$.}
\label{figfpm}
\end{center}
\end{figure}

Note that the critical configuration type II (IV) appears when 
${\widetilde\Delta}^2_+=0$
(${\widetilde\Delta}^2_-=0$).
From Eq.~(\ref{Discussc}), the condition ${\widetilde\Delta}^2_\pm=0$
is rephrased as 
\begin{eqnarray}
f^{\pm} (\xi)=-8\left({\widetilde{\delta r_s}\over \eta}\right)^2,
\label{fpm}
\end{eqnarray}
where we defined
\begin{eqnarray}
f^{\pm} (\xi)=\xi^4-20\xi^2-8\pm\xi^3\sqrt{8+\xi^2}\pm8\xi\sqrt{8+\xi^2},
\label{deffpm}
\end{eqnarray}
and $\xi=\widetilde {\delta y_{10}}/2\eta$. 
Fig.~\ref{figfpm} plots $f^{+} (\xi)$ and $f^{-} (\xi)$ as a 
function of $\xi$. 
Note that $f^{\pm} (0)=-8$.
From Fig.~\ref{figfpm}, one finds $f^{+} (\xi)\geq f^-(\xi)$. This  
means that ${\widetilde\Delta}^2_+\geq {\widetilde\Delta}^2_-$,
which can be proved explicitly. This also means that the 
configuration type always changes as 
${\rm V}\rightarrow {\rm IV}\rightarrow{\rm III}\rightarrow 
{\rm II}\rightarrow{\rm I}$, as ${\widetilde{\delta y_{10}}
/\eta}$ changes from {\it infinity} to $0$.

We can easily find the critical condition that the type II 
appears in an analytic way, as follows.
The condition is ${\widetilde\Delta}^2_+=0$, i.e., Eq.~(\ref{fpm})
of the $+$ sign, which has a solution around $\xi\sim1$ 
for $\eta\sim\widetilde{\delta y_{10}}$. By expanding $f^{+}(\xi)$ 
around $\xi=1$, we have
\begin{equation}
f^{+}(\xi)\simeq\frac{256}{27}(\xi-1)^3+\mathcal{O}(\xi-1)^4.
\label{Discussg}
\end{equation}
With this approximation, ${\widetilde\Delta}^2_+=0$ yields 
\begin{equation}
{\widetilde{\delta y_{10}}\over 2\eta}=1-\frac{3}{2}\left(\frac{\widetilde{\delta r_s}}{2\eta}\right)^{2/3}.
\label{boundaryap}
\end{equation}

\begin{figure}[htbp]
\begin{center}
\includegraphics[width=120mm, height=120mm]{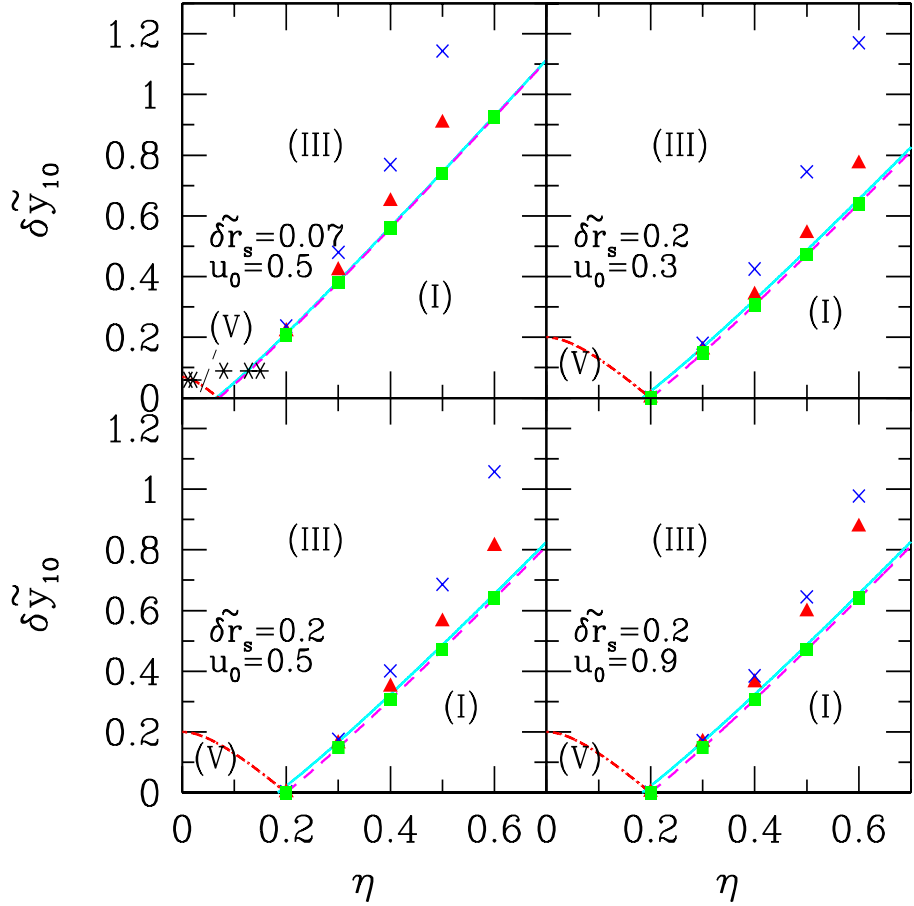}
\end{center}
\caption{Boundary of the type I, III and V on the 
$\eta-\widetilde{\delta y_{10}}$ plane. 
The dashed curves satisfy $\widetilde\Delta^2_+=0$, 
the dot-dashed curves satisfy $\widetilde\Delta^2_-=0$, 
which are obtained by solving Eq.~(\ref{fpm}) of the sign $+$ and 
$-$, respectively, 
while the solid curves plot Eq.~(\ref{boundaryap}).
In order to show the validity of the approximate approach, 
we plot the points of the critical configuration 
type II, the crosses are the exact approach (a), 
the triangles are the perturbative approach (b), 
and the squares are the approximate approach(c), 
respectively. The five asterisks in the upper left panel 
correspond to the value of $\eta$ and $\widetilde{\delta y_{10}}$
of Figs.~1 $\sim$ 5 from the right to left, respectively.}
\label{fig:boundary}
\end{figure}

Fig.~\ref{fig:boundary} is the diagram to show where the typical 
configurations appear on the $\eta-\widetilde{\delta y_{10}}$ plane. 
On the boundary between I and III, the critical configuration 
type II appears, while the type IV appears on the boundary between 
III and V. 
The dashed curves 
satisfy $\widetilde\Delta^2_+=0$, the dot-dashed  
curves satisfy $\widetilde\Delta^2_-=0$, which are obtained by
solving Eq.~(\ref{fpm}) of the sign $+$ and $-$, respectively. 
The solid curves plot Eq.~(\ref{boundaryap}).
The agreement between the dashed curve
and the solid curve means the validity of the approximate formula
of (\ref{boundaryap}).
Here, we adopted $\widetilde{\delta r_s}=0.07$, $u_0=0.5$ 
(upper-left panel) and $\widetilde{\delta r_s}=0.2$, $u_0=0.3$ 
(upper-right panel), respectively.
The lower left (right) panel assumes $u_0=0.5$ ($u_0=0.9$). 

The critical boundary of Fig.~\ref{fig:boundary} is obtained 
on the basis of the approximate approach of the lowest order 
expansion in terms of $\eta$. 
The exact critical value should be found by using figures like 
Fig.~\ref{figthetaeta}. 
The points in Fig.~\ref{fig:boundary} 
are obtained by making figure like Fig.~\ref{figthetaeta}. 
The cross, the triangle, and the square are the results with 
the approach (a), (b) and (c), respectively. 
Thus, the critical boundary is estimated lower when 
we use the approximate approach, as shown in each panel of 
Fig.~\ref{fig:boundary}. 
However, this figure also demonstrates that the approximate 
approach is quite good as long as $\eta\simlt0.3$.

The condition that the critical configuration
II appears is $\widetilde\Delta^2_-=0$. We may write the condition as
\begin{equation}
\widetilde{\delta y_{10}}+\eta\simlt \widetilde{\delta r_s}.
\end{equation}
%

\subsection{Relation with caustics and critical line}
We consider the condition that the circumference of the source comes 
in contact with the caustics. 
The intersection point is obtained by substituting Eqs.~(\ref{caua}) and
(\ref{caub}) into Eqs.~(\ref{sourcea}) and (\ref{sourceb}), 
which yields
\begin{eqnarray}
&&\left({\partial^2 \widetilde{\delta\phi}\over \partial
 \theta^2}\Bigr|_{x=1}\right)^2+\left({\partial \widetilde{\delta\phi}\over \partial
 \theta}\Bigr|_{x=1}\right)^2+\widetilde{\delta y_{10}}^2
 \nonumber \\&&~~~~~~~
-2\widetilde{\delta y_{10}}\left({\partial^2 \widetilde{\delta\phi}\over \partial
 \theta^2}\Bigr|_{x=1}\cos\theta+{\partial \widetilde{\delta\phi}\over \partial
 \theta}\Bigr|_{x=1}\sin\theta\right)=\widetilde{\delta r_s}^2,
\label{intersection}
\end{eqnarray}
where we used $\widetilde{\delta y_{20}}=0$. 
Within the approximate approach (c), we use
Eqs.~(\ref{dthetadeltaphiap}) and (\ref{appendoox}), 
which we substitute into Eq.~(\ref{intersection}), then
\begin{eqnarray}
g(\theta)\equiv 12\eta^2\cos^4\theta-4\eta\widetilde{\delta
 y_{10}}\cos^3\theta-12\eta^2\cos^2\theta+4\eta^2+\widetilde{\delta y_{10}}^2-\widetilde{\delta r_s}^2=0.
\nonumber
\\
\label{intetsec}
\end{eqnarray}
When the circumference of the source comes in contact with the caustics, 
the solution of Eq.~(\ref{intetsec}) has only one solution.
This condition is 
\begin{equation}
g(\theta_c)=0,
\label{intcon}
\end{equation}
where $\theta_c$ satisfies
\begin{equation}
{d g(\theta)\over d\theta}\bigg|_{\theta=\theta_c}=0.
\label{gceq}
\end{equation}
The solution of Eq.~(\ref{gceq}) is
\begin{eqnarray}
\theta_{c\pm}=\cos^{-1}\left[{1\over 8\eta} \left(
\widetilde{\delta y_{10}}\pm \sqrt{32\eta^2+\widetilde{\delta y_{10}}^2}
\right)\right],
\label{cauc}
\end{eqnarray}
then, Eq.(\ref{intcon}) gives 
\begin{eqnarray}
g(\theta_{c\pm})=&&\frac{1}{128\eta^2}\Bigl[-\delta y_{10}^4
+80\eta^2\delta y_{10}^2+128\eta^4\pm\delta
y_{10}^3\sqrt{32\eta^2+\delta y_{10}^2}
\nonumber \\&&~~
\pm32\eta^2\delta y_{10}\sqrt{32\eta^2+\delta
y_{10}^2}+128\eta^2\delta r_s^2\Bigr]=0.
\end{eqnarray}
This is the condition that the circumference of the source comes in 
contact with the  caustics. Note that
$\theta_{c\pm}=\theta_{min\pm}$, and $g(\theta_{c\pm})=\widetilde{\Delta}_{\pm}^{2}$. 
This means that the
critical configuration type II (plus sign) and type IV (minus sign) 
appear when the circumference of the source comes in 
contact with the  caustics.

This condition can be transformed to the following relation 
between the critical line and the image. From Eq.~(\ref{dreqap}), 
the central line of the image can be defined by
\begin{eqnarray}
  &&\widetilde {\delta x}={1\over (1-\partial_{\tilde x}^2 
\widetilde{\phi_0}(\widetilde x))}\Bigl[
\partial_{\widetilde x}\widetilde{\delta\phi}(\widetilde x,\theta)+\widetilde{\delta y_{10}}
\cos\theta\Bigr]\Bigr|_{\widetilde x=1}.
\label{dreqapp}
\end{eqnarray}
The critical line is defined by Eq.~(\ref{clap}), then 
an intersection point of the critical line and 
the central line of the image satisfies
\begin{equation}
\widetilde{\delta y_{10}}\cos\theta={\partial^2 \widetilde{\delta\phi}\over \partial
 \theta^2}\Bigr|_{x=1}.
\end{equation}
Within the approximate approach (c), using Eq.~(\ref{appendoox}),
this condition gives
\begin{equation}
\widetilde{\delta y_{10}}\cos\theta=2\eta\cos 2\theta,
\end{equation}
which can be solved easily, 
\begin{eqnarray}
\cos\theta_{\rm cri\pm}={1\over 8\eta} \left(
\widetilde{\delta y_{10}}\pm \sqrt{32\eta^2+\widetilde{\delta y_{10}}^2}
\right).
\end{eqnarray}
Note that $\theta_{\rm cri\pm}=\theta_{\rm min\pm}(=\theta_{c\pm})$. 

The above behaviour of the critical configuration is obtained using 
the approximate approach (c), but holds in the 
exact approach in a similar way. 
Indeed, these critical behaviour can be seen in Figs.~\ref{figII} 
and \ref{figIV}. 
These facts also guarantee the usefulness of the approximate 
approach to investigate the lensing phenomena in a simple analytic 
way.

\subsection{Application of approximate approach}

In this subsection, let us summarise a few useful consequences, 
which are obtained using the approximate approach in an analytic 
manner.

First, we consider the width of lensed images. Within the perturbative
approach, the width of the image is 
\begin{eqnarray}
\widetilde{\delta x_+}-\widetilde{\delta x_-}&=&{2\over
 (1-\partial_{\tilde x}^2\phi_0(\widetilde x))}
\sqrt{\Delta^2(\widetilde x,\theta)}\biggr|_{\widetilde x=1}.
\end{eqnarray}
From Eq.~(\ref{Discusse}), the maximum width at $\theta=\theta_{\rm max}$ is 
\begin{eqnarray}
(\widetilde{\delta x_+}-\widetilde{\delta x_-})_{\rm max}&=&
 {2 \widetilde{\delta r_s} \over (1-\partial_{\tilde  x}^2
 \phi_0(\widetilde x))}.
\end{eqnarray}
From Eq.~(\ref{Discussf}), the same maximum width appears at 
$\theta=0$ and $\pi$.

Second, let us consider the angular size of the arc, which can be
obtained by solving $\widetilde{\Delta}^2(\theta)=0$ 
under the condition
$\widetilde{\Delta}^2_+>0$, because $\widetilde{\Delta}^2(\theta)\geq0$ is 
necessary for the appearance of the image. 
Using Eq.~(\ref{Discussa}), $\widetilde{\Delta}^2(\widetilde x=1,\theta)=0$  
reduces to
\begin{eqnarray}
\widetilde{\delta r_s}^2-(\eta\sin2\theta
-\widetilde{\delta y_{10}}\sin\theta)^2=0,
\end{eqnarray}
which can be solved easily with a suitable method. 

Third, we focus on the magnification of the lensed image. The magnification 
factor of an extended source is written (e.g.,\cite{SEF}),
\begin{eqnarray}
\mu_e={\int\mathrm{d}^2yI(\vec{y})\mu(\vec{y})
\over \int\mathrm{d}^2yI(\vec{y})},
\end{eqnarray}
where $I(\vec y)$ is the surface brightness at $\vec y$.
In the case the surface brightness is a constant, i.e.,
$I(\vec{y})=I_0$,  we may write
\begin{equation}
\mu_e={\int_{|\widetilde{\vec y}|<\delta r_s}\mathrm{d}^2
\widetilde {\vec y}\mu(\widetilde{\vec{y}})
\over \pi \widetilde{\delta r_s}^2}.
\end{equation}
From the definition of the magnification $\mu$, we have 
\begin{eqnarray}
\mu_e=\int d\theta {dA\over d\theta},
\end{eqnarray}
where we defined
\begin{equation}
{dA\over d\theta}=
{1\over \pi \widetilde{\delta r_s}^2}
\int^{1+\widetilde{\delta x_+}}_{1+\widetilde{\delta x_-}}\widetilde{r'}
\mathrm{d}\widetilde{
r'}
={1\over \pi \widetilde{\delta r_s}^2}
(\widetilde{\delta x_+}-\widetilde{\delta x_-}).
\end{equation}
Within the perturbative approach, we have
\begin{eqnarray}
{dA\over d\theta}
&=&{1\over \pi \widetilde{\delta r_s}^2}
 {2\sqrt{\Delta^2(\widetilde x,\theta)}\over (1-\partial_{\tilde x}^2
 \phi_0(\widetilde x))}\biggr|_{\widetilde x=1}.
\label{toda}
\end{eqnarray}
In the approximate method (c),
we may substitute the expression (\ref{Discussa}) into Eq.~(\ref{toda}).

\begin{figure}[htbp]
\begin{center}
\includegraphics[width=120mm, height=120mm]{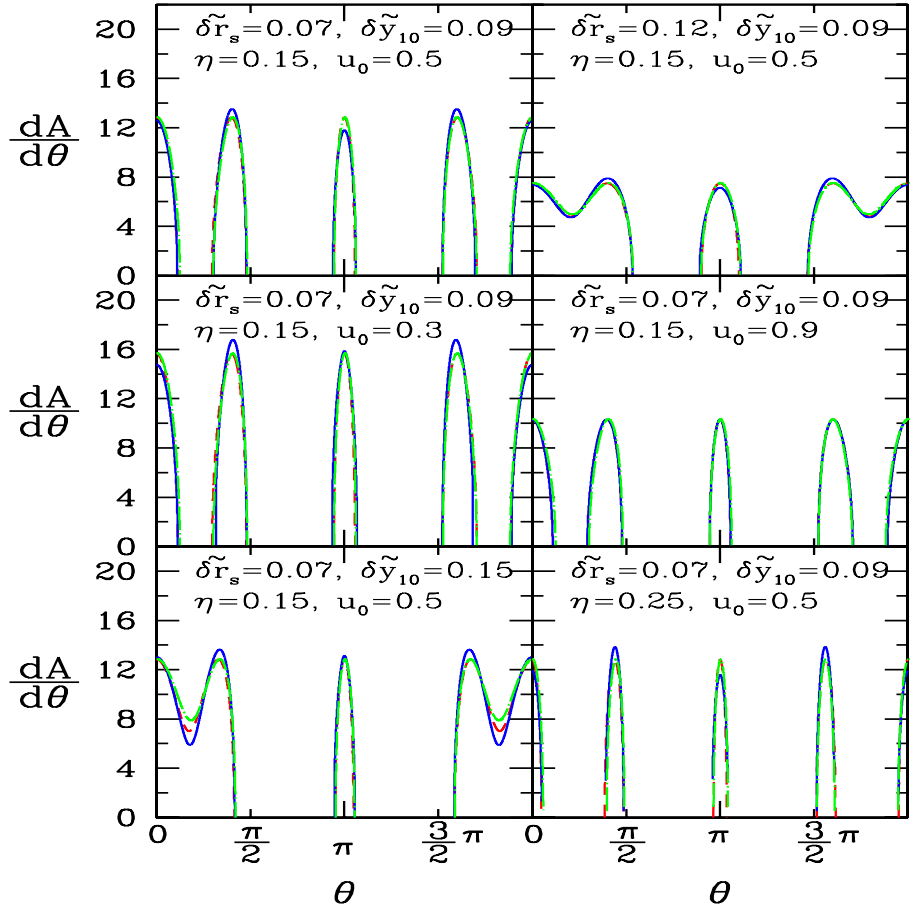}
\end{center}
\caption{$dA/d\theta$ as a function of $\theta$.
A different set of the parameters $\widetilde{\delta r_s}$, 
$\widetilde{\delta y_{10}}$, $\eta$ and $u_0$ are adopted for 
each panel as shown therein. The solid curves are the exact approach (a), 
the dashed curves are the perturbative approach (b), and the 
long dash-dotted curves are the approximate approach (c). } 
\label{fig:dmudt}
\end{figure}

Fig.~\ref{fig:dmudt} shows $dA/d\theta$ as a function of $\theta$. 
In each panel, a different set of the parameters $\widetilde{\delta r_s}$, 
$\widetilde{\delta y_{10}}$, $\eta$ and $u_0$ is adopted, as shown therein. 
The solid curves show the exact approach (a), 
the dashed curves show the perturbative approach (b), and the 
long dash-dotted curves show the approximate approach (c). 
The top left panel adopts the same parameters as those of
Fig.~\ref{figI}. 
In the top right panel, $\widetilde{\delta r_s}$ is increased 
compared with the top left panel. 
This increase of $\widetilde{\delta r_s}$ changes the four lensed 
images to one arc and the other separated image. 
In the middle left (right) panel, compared with the top left panel, 
$u_0$ is decreased (increased), 
by which the amplification is increased (decreased).
In the lower left panel, compared with the top left panel,
$\widetilde{\delta y_{10}}$ is increased, by which  
the four separate images change to one arc and the other separated image. 
In the lower right panel, $\eta$ is increased.

From Eqs.~(\ref{Discusse}) and (\ref{Discussf}), we have 
$\Delta^2(\widetilde x=1,\theta)=\widetilde{\delta r_s}^2$
for $\theta=0,~\theta_{\rm max},~\pi$, where the width of the image 
becomes maximum. Then, the maximum value of $dA/d\theta$ is
\begin{equation}
{dA\over d\theta}={2\over \pi\widetilde{\delta r_s}(1-\partial_{\tilde x}^2
 \phi_0(\widetilde x))|_{\widetilde x=1}}.
\end{equation}
Thus, $dA/d\theta$ takes the same value 
at $\theta=0,~\theta_{\rm max},~\pi$, which can be also seen 
from Fig.~\ref{fig:dmudt}.

\subsection{Point source limit}
Finally, in this section, we consider the limit of a point source, 
which is given by imposing $\widetilde{\delta r_s}=0$.
In this limit, Eq.~(\ref{Delta2ap}) yields
\begin{eqnarray}
  &&\widetilde\Delta^2(\widetilde x,\theta)=
-\Bigl(
  {1\over \widetilde x}\partial_\theta\widetilde{\delta \phi}
(\widetilde x,\theta)
  -\widetilde{\delta y_{10}}\sin\theta
  +\widetilde{\delta y_{20}}\cos\theta \Bigr)^2.
\end{eqnarray}
For the existence of the solution, we must have  
$\widetilde \Delta(\widetilde x,\theta)=0$. 
In the approximate approach (c), with the use of 
Eq.~(\ref{dthetadeltaphiap}),
$\widetilde \Delta(\widetilde x=1,\theta)=0$ yields
\begin{equation}
\eta\sin2\theta-\widetilde\delta y_{10}\sin\theta
+\widetilde{\delta y_{20}}\cos\theta=0.
\label{pointequationb}
\end{equation}
This equation is equivalent to 
\begin{equation}
\eta\sin2\theta-
\sqrt{\widetilde{\delta y_{10}}^2+\widetilde{\delta y_{20}}^2}
\sin(\theta-\theta_0)=0,
\label{inclinationequation}
\end{equation}
where $\theta_0=\arctan(\widetilde\delta y_{20}/\widetilde\delta y_{10})$ 
is regarded as an inclination angle when choosing 
the coordinate so that the point source is located on the 
$\widetilde{y_{1}}$-axis.
It is easy to solve Eq.~(\ref{inclinationequation}) in a numerical method.
When a solution of Eq.~(\ref{inclinationequation}) is found, which 
we denote by $\theta_p$, from Eqs.~(\ref{dreqap}) and 
(\ref{perturbativeJ}), we obtain the image position and 
the magnification factor, 
\begin{equation}
\widetilde {\delta x}={1\over (1-\partial_{\tilde x}^2 
\widetilde{\phi_0}(\widetilde x))}\Bigl[
\partial_{\widetilde x}\widetilde{\delta\phi}(\widetilde
x,\theta)+\widetilde{\delta y_{10}}\cos\theta\Bigr]\Biggr|_{\widetilde x=1,\theta=\theta_{p}}
\label{imagaggg}
\end{equation}
and
\begin{equation}
\mu\simeq\biggl[(1-\partial_{\tilde x}^2 
\widetilde{\phi_0}(\widetilde x))\left(\widetilde{\delta y_{10}}\cos\theta-2\eta\cos2\theta\right)\biggr]^{-1}\biggr|_{\widetilde x=1,\theta=\theta_{p}},
\label{magniffff}
\end{equation}
respectively. 

For simplicity, we here consider the case $\widetilde{\delta y_{20}}=0$,
that is, the inclination angle is zero. In this case, 
Eqs. (\ref{imagaggg}) and (\ref{magniffff}) yield
simple analytic expressions, as follows.
Eq. (\ref{pointequationb}) reduces to 
\begin{equation}
\eta\sin2\theta-\widetilde\delta y_{10}\sin\theta=0,
\label{pointequation}
\end{equation}
which gives the following solution to represent the angular 
position of the image,
\begin{equation}
\cos\theta_{p}=\frac{\widetilde{\delta y_{10}}}{2\eta}
\label{pointsolution}
\end{equation}
and
\begin{equation}
\sin\theta_{p}=0.
\label{pointsolutionsin}
\end{equation}
Eq.~(\ref{imagaggg}) gives the radial position of the image. 
For the solution of Eq.(\ref{pointsolution}), we have
\begin{equation}
\widetilde {\delta x}={1\over (1-\partial_{\tilde x}^2 
\widetilde{\phi_0}(\widetilde x))}\Bigl[
\partial_{\widetilde x}\widetilde{\delta\phi}(\widetilde
x,\theta)+\frac{\widetilde{\delta
y_{10}}^2}{2\eta}\Bigr]\Bigr|_{\widetilde x=1,\theta=\theta_{p}}.
\label{radsolutionp}
\end{equation}
By substituting Eq.~(\ref{radsolutionp}) into
Eq.~(\ref{magniffff}), we have the magnification factor,
\begin{eqnarray}
\mu=\biggl[(1-\partial_{\tilde x}^2 
\widetilde{\phi_0}(\widetilde x))\left(\frac{\widetilde{\delta
y_{10}}^2}{2\eta}-{\partial^2\widetilde{\delta\phi}\over \partial\theta^2}
\right)\biggr]^{-1}\biggr|_{\widetilde x=1,\theta=\theta_{p}}.
\end{eqnarray}
In the approximate approach (c), using Eqs.~(\ref{appendoox}) 
and (\ref{pointsolution}), 
this magnification factor is expressed 
\begin{equation}
\mu\simeq\biggl[2\eta(1-\partial_{\tilde x}^2 
\widetilde{\phi_0}(\widetilde x))\left(1-\left(\frac{\widetilde{\delta
y_{10}}}{2\eta}\right)^2\right)\biggr]^{-1}\biggr|_{\widetilde x=1}.
\end{equation}
In the same way, for the solution of Eq.~(\ref{pointsolutionsin}), we 
obtain the radial position of the image, 
\begin{equation}
\widetilde {\delta x}={1\over (1-\partial_{\tilde x}^2 
\widetilde{\phi_0}(\widetilde x))}\biggl[
\partial_{\widetilde x}\widetilde{\delta\phi}(\widetilde
x,\theta)\pm\widetilde{\delta y_{10}}\biggr]\Bigr|_{\widetilde x=1,\theta=\theta_{p}},
\label{radsolutionpsin}
\end{equation}
and the magnification factor,
\begin{equation}
\mu\simeq\biggl[2\eta(1-\partial_{\tilde x}^2 
\widetilde{\phi_0}(\widetilde x))\left(-1\pm\frac{\widetilde{\delta
y_{10}}}{2\eta}\right)\biggr]^{-1}\biggr|_{\widetilde x=1},
\end{equation}
where $\partial_{\tilde x}^2 
\widetilde{\phi_0}(\widetilde x)\bigr|_{\widetilde x=1}$
is given by Eq.~(\ref{dxxphi}), where the sign $+$ and $-$ correspond 
to the solution $\theta_p=0$ and $\theta_p=\pi$, respectively, 
from Eq.~(\ref{pointsolutionsin}).


\section{Summary and Conclusions}
We studied the perturbative approach to the strong lensing system,
which was extended by Alard, in both the analytic and numerical 
manners. We investigated the validity of the perturbative approach
by comparing with the exact approach on the basis of the numerical 
method, focusing on the shape of the image, the magnification, the
caustics, and the critical line. The perturbative approach works
well in the case when the ellipticity of the lens potential $\eta$
is small and the configuration of the source is close to the that
of an Einstein ring. 
At a quantitative level, the perturbative approach is valid at the
$10$ percent level for $\widetilde{\delta y_{10}}\simlt0.2$ and
$\eta\simlt 0.3$. We also demonstrated that the lowest-order 
expansion in terms of $\eta$ also works well, which enables
us to investigate the lensing system in an analytic way.

We investigated the critical behaviour of the lensed images, 
by demonstrating the phase diagram of the different configurations
of four separated images (type I), 
an arc and one separated image (type III), and
one connected ring image (type V). 
The critical configuration of type II appears during the transition
from the type I to the type III, while the type IV appears during the 
transition from the type III to the type V.
We investigated how the critical behaviour depends on 
the lens ellipticity, the source position and the source radius.
We also demonstrated how the appearance of the critical configuration 
III and V is related to the condition 
between the source configuration and the caustics. 
The condition of the critical configuration 
was investigated in an analytic manner using the 
lowest-order expansion of the ellipticity $\eta$ of the elliptical lens
potential.

The perturbative approach with the lowest-order expansion with respect 
to $\eta$ is useful to find the simple formulas which characterises
the lensing system in an analytic manner. 
We derived the analytical formulas of the arc width and the 
magnification factor. 
In the point source limit, the simple formulas for 
the image position and magnification factor were obtained.
These formulas can be easily solved, which gave the simple 
analytic expressions in the absence from the inclination angle.
These results will be useful to understand the gravitational 
lensing phenomena. 

In a realistic situation in reconstructing a gravitational lensing 
system, substructures in the lens might have to be taken into account. 
In the reference,\cite{Alard08} Alard considered how a substructure
affects a lensed image in the perturbative approach. 
Even a substructure with small mass could make a change in the caustics 
and the lensed image drastically. It is an interesting problem 
how one can determine the gravitational lens potential 
including substructures simultaneously. Here, there is
potentially a lot of room for improvement.\cite{Alard09} 
This issue is outside the scope of the present paper, but need to 
be elaborated for a precise reconstruction of a gravitational 
lens system.

\section*{Acknowledgements}
This work is supported by Japan Society for Promotion
of Science (JSPS) Grants-in-Aid
for Scientific Research (Nos.~21540270,~21244033).
This work is also supported by JSPS 
Core-to-Core Program ``International Research 
Network for Dark Energy''.
We thank M. Meneghetti and M. Oguri for useful discussions. 

\begin{appendix}
\section{Formulas in the exact approach}
In this appendix, we summarize formulas of the 
lens equation in the exact approach (a) without any approximation. 
We consider the lensed image of the circumference of the 
circular source, whose center is located at $(\widetilde{\delta y_{10}},
\widetilde{\delta y_{20}})$. The source's radius is $\widetilde{\delta r_s}$. 
The circumference of the circular source is parameterized as
\begin{eqnarray}
&&\widetilde{y_1}=\widetilde{\delta y_{10}}+\widetilde{\delta r_s} \cos\varphi,
\\
&&\widetilde{y_2}=\widetilde{\delta y_{20}}+\widetilde{\delta r_s} \cos\varphi
\end{eqnarray}
with the parameter $\varphi$ in the range $0\leq \varphi\leq 2\pi$.
As the image is parameterized as
\begin{eqnarray}
&&\widetilde{x_1} = \widetilde{x} \cos\theta,
\\
&&\widetilde{x_2} = \widetilde{x} \sin\theta,
\end{eqnarray}
the lens equation is 
\begin{eqnarray}
  &&\widetilde{\delta y_{10}}+\widetilde{\delta r_s} \cos\varphi=\widetilde{x}\cos\theta
  -{\cos\theta}{\partial\over \partial \widetilde{x}} \widetilde{\phi}(\widetilde{x},\theta)
+{\sin\theta\over \widetilde{x}}{\partial\over \partial \theta} \widetilde{\phi}(\widetilde{x},\theta),
\label{lenseqexx}
\\
  &&\widetilde{\delta y_{20}}+\widetilde{\delta r_s} \sin\varphi=\widetilde{x}\sin\theta
  -{\sin\theta}{\partial\over \partial \widetilde{x}} \widetilde{\phi}(\widetilde{x},\theta)
   -{\cos\theta\over \widetilde{x}}{\partial\over \partial \theta} \widetilde{\phi}(\widetilde{x},\theta),
\label{lenseqexy}
\end{eqnarray}
which yield
\begin{eqnarray}
  &&\widetilde{x}=\widetilde{\delta y_{10}} \cos\theta + \widetilde{\delta y_{20}} \sin\theta 
  +{\partial\over\partial \widetilde{x}}\widetilde{\phi}(\widetilde{x},\theta)
    \nonumber
\\&&~~~
  \pm \sqrt{\widetilde{\delta r_s}^2 -\left({1\over \widetilde{x}}{\partial \over \partial \theta}
  \widetilde{\phi}(\widetilde{x},\theta)-\widetilde{\delta y_{10}} \sin\theta+\widetilde{\delta y_{20}}\cos\theta\right)^2},
\label{exactimage}
\end{eqnarray}
and 
\begin{eqnarray}
 &&\tan(\theta-\varphi)=
\left[{1\over \widetilde{x}}{\partial \over \partial \theta}\widetilde{\phi}(\widetilde{x},\theta)
 -\widetilde{\delta y_{10}}\sin\theta+\widetilde{\delta y_{20}}\cos\theta \right]
    \nonumber
\\&&~~~~~~~~~~~~~~~\times
\left[\widetilde{x}-{\partial \over \partial \widetilde{x}}\widetilde{\phi}(\widetilde{x},\theta)
 -\widetilde{\delta y_{10}}\cos\theta-\widetilde{\delta y_{20}}\sin\theta\right]^{-1}.
\end{eqnarray}

For the elliptical NFW lens model adopted in this paper, the potential  
is written,
\begin{eqnarray}
 \widetilde\phi={4\kappa_s\over u_0^2}{1\over \widetilde x}
\biggl(\log{ \Xi\over 2}
-2{\rm arctanh}^2
\sqrt{{1-\Xi}\over 1+\Xi}\biggr),
\label{ENFWE}
\end{eqnarray}
where we defined $\Xi=u_0\widetilde x\sqrt{1-\eta \cos
2\theta}$. Here, the expression of the case $\Xi<1$ 
is presented. Eq.~(\ref{ENFWE}) gives
\begin{eqnarray}
 &&{\partial\widetilde\phi\over \partial \tilde{x}}
 ={4\kappa_s\over u_0^2}{1\over \widetilde x}\biggl(
\log{ \Xi\over 2}+{2\over \sqrt{1-\Xi^2}}
{\rm arctanh}\sqrt{1-\Xi \over 1+\Xi}
\biggr),
\label{explicita}
\\
 &&{\partial\widetilde\phi\over \partial \theta}
 ={4\kappa_s\over u_0^2}{\eta\sin2\theta\over (1-\eta\cos2\theta)}\biggl(
\log{ \Xi\over 2}+{2\over \sqrt{1-\Xi^2}}
{\rm arctanh}\sqrt{1-\Xi \over 1+\Xi}
\biggr),
\label{explicitb}
\\
 &&{\partial\widetilde\phi\over \partial \theta \partial \tilde{x}}
 ={4\kappa_s\over u_0^2}u_0^2\eta\tilde{x}\sin2\theta
\biggl({1\over \Xi^2-1}
\nonumber\\
&&~~~~~~+{1\over (1-\Xi^2)^{3/2} }
{\rm arctanh}\sqrt{1-\Xi \over 1+\Xi}
\biggr),
\label{explicitc}
\\
 &&{\partial\widetilde\phi\over \partial \tilde{x}^2}
 ={4\kappa_s\over u_0^2}{1 \over \widetilde x^2}\biggl(
-\log{ \Xi\over 2}-{\Xi^2\over 1-\Xi^2
}\nonumber\\
&&~~~~~~+{2(-1+2\Xi^2)\over \sqrt{1-\Xi^2}}
{\rm arctanh}\sqrt{1-\Xi \over 1+\Xi}
\biggr),
\label{explicitd}
\\
 &&{\partial\widetilde\phi\over \partial \tilde{\theta}^2}
 ={4\kappa_s\over u_0^2}\biggl(
{2\eta(\cos2\theta-\eta)\over (1-\eta\cos2\theta)^2}\log{\Xi\over 2}
-{u_0^2\widetilde x^2\eta^2\sin^2 2\theta\over
(1-\eta\cos2\theta)(1-\Xi^2)}\nonumber\\
&&~~~~~~{4\eta(1-\Xi^2)(\cos2\theta-\eta)+2\eta^2\Xi^2\sin^2
 2\theta\over (1-\eta\cos2\theta)^2(1-\Xi^2)^{3/2}}{\rm
 arctanh}\sqrt{1-\Xi \over 1+\Xi}\biggr).
\label{explicite}
\end{eqnarray}
The case $\Xi<1$ is given by the analytic continuation. 

\newpage
\section{Formulas in the perturbative approach}
We here summarise useful formulas in the perturbative approach (b).
For the elliptical NFW lens potential, we have
\begin{eqnarray}
&&{\partial^2 \over \partial {\widetilde x}^2} \widetilde \phi_0
\Bigr|_{\widetilde x=1}={4\kappa_s\over u_0^2} \biggl(
-\log{u_0\over 2}-{u_0^2\over 1-u_0^2}
\nonumber\\
&&~~~~~~+2{-1+2u_0^2\over (1-u_0^2)^{3/2}}{\rm arctanh}
\sqrt{1-u_0 \over 1+u_0}\biggr),
\label{dxxphi}
\\
&& 
{\partial \over \partial {\widetilde x}} \widetilde{\delta\phi}
\Bigr|_{\widetilde x=1}={4\kappa_s\over u_0^2} \biggl(
\log\Upsilon-{2\over \sqrt{1-u_0^2}}
{\rm arctanh}\sqrt{1-u_0 \over 1+u_0}
\nonumber\\
&&~~~~~~~~~+{2\over \sqrt{1-u_0^2 \Upsilon^2}}{\rm arctanh}
\sqrt{{1-u_0\Upsilon}\over 1+u_0\Upsilon}\biggr),
\label{dxdeltaphi}
\\
&& 
{\partial \over \partial {\theta}} \widetilde{\delta\phi}
\Bigr|_{\widetilde x=1}={4\kappa_s\over u_0^2} {\eta \sin2\theta \over \Upsilon^2}
\biggl(\log{ u_0\Upsilon\over 2}
\nonumber\\
&&~~~~~~~~~+{2\over \sqrt{1-u_0^2 \Upsilon^2}}{\rm arctanh}
\sqrt{{1-u_0\Upsilon}\over 1+u_0\Upsilon}\biggr),
\label{dthetadeltaphi}
\\
&& 
{\partial^2 \over \partial {\theta}^2} \widetilde{\delta\phi}
\Bigr|_{\widetilde x=1}={4\kappa_s\over u_0^2} {1 \over \Upsilon^4}
\biggl(2(1-\Upsilon^2-\eta^2)\log{ u_0\Upsilon\over 2}
\nonumber\\
&&~~~~~~~~~+{2\eta^2 u_0^2 \Upsilon^2\sin 2\theta
\over \sqrt{1-u_0^2 \Upsilon^2}^{3}}{\rm arctanh}
\sqrt{{1-u_0\Upsilon}\over 1+u_0\Upsilon}
\nonumber\\
&&~~~~~~~~~+{4(1-\Upsilon^2-\eta^2)
\over \sqrt{1-u_0^2 \Upsilon^2}}{\rm arctanh}
\sqrt{{1-u_0\Upsilon}\over 1+u_0\Upsilon}\biggr)
\nonumber\\
&&~~~~~~~~~-{\eta^2 u_0^2 \Upsilon^2\sin^2 2\theta
\over 1-u_0^2 \Upsilon^2}
\biggr),
\label{dttdeltaphi}
\end{eqnarray}
where we defined $\Upsilon=\sqrt{1-\eta \cos 2\theta}$.

\section{Formulas in the approximate approach}
In the approximate approach (c), the potential is approximated
at the lowest order of expansion with respect to $\eta$. 
The following formulas are useful. By expanding the potential 
$\widetilde{\delta\phi}$, Eq.~(\ref{deltaphiNFW}), we have
\begin{eqnarray}
&&\widetilde{\delta\phi}(\widetilde x,\theta)\simeq 
{4\kappa_s\over u_0^2}\eta\cos 2\theta
\biggl(-{1\over 2} \log {u_0 \widetilde x\over 2}
-{1\over (1-u_0^2)^{1/2} }
{\rm arctanh}\sqrt{1-u_0\widetilde x\over 1+u_0\widetilde x}\biggr) ,
\nonumber
\\
\label{appendox}
\end{eqnarray}
at the lowest order of $\eta$. Similarly, 
Eqs.~(\ref{dxdeltaphi})$\sim$(\ref{dttdeltaphi}) lead to
\begin{eqnarray}
&&{\partial \over \partial {\widetilde x}} \widetilde{\delta\phi}
\Bigr|_{\widetilde x=1}\simeq {4\kappa_s\over u_0^2} \eta \cos 2\theta 
\biggl({u_0^2 \over 2(1-u_0^2)}
-{u_0^2\over (1-u_0^2)^{3/2} }
{\rm arctanh}\sqrt{1-u_0 \over 1+u_0}\biggr),
\nonumber
\\
\label{dxdeltaphiap}
\\
&& 
{\partial \over \partial {\theta}} \widetilde{\delta\phi}
\Bigr|_{\widetilde x=1}
\simeq
{\eta \sin2\theta},
\label{dthetadeltaphiap}
\\
&& 
{\partial^2 \over \partial {\theta}^2} \widetilde{\delta\phi}
\Bigr|_{\widetilde x=1}
\simeq
{2\eta \cos 2\theta}.
\label{appendoox}
\end{eqnarray}


\end{appendix}


\end{document}